\pdfoutput=1
\documentclass[12pt,a4paper]{article}

\usepackage{ifthen} 
\newboolean{pdflatex}
\setboolean{pdflatex}{true} 

\newboolean{articletitles}
\setboolean{articletitles}{true} 

\newboolean{uprightparticles}
\setboolean{uprightparticles}{false} 

\def\paperauthors{LHCb collaboration} 
\def\paperasciititle{First observation of the Bs0->Lambda\_c+ Lambda\_c- and evidence for the B0->Lambda\_c+ Lambda\_c- decay} 
\def\papertitle{First observation of the \decay{\Bsb}{\Lc\Lcbar} decay and evidence for the \decay{\Bdb}{\Lc\Lcbar} decay} 
\def\paperkeywords{{High Energy Physics}, {LHCb}} 
\def\papercopyright{\the\year\ CERN for the benefit of the LHCb collaboration} 
\def\paperlicence{CC BY 4.0 licence}
\def\paperlicenceurl{https://creativecommons.org/licenses/by/4.0/}

\newif\ifEnableSectionTOCLinks
\EnableSectionTOCLinksfalse 

\usepackage[top=1in, bottom=1.25in, left=1in, right=1in]{geometry}

\usepackage{microtype}
\usepackage{lineno}  
\usepackage{xspace} 
\usepackage{caption} 

\usepackage{graphicx}  
\usepackage{color}
\usepackage{colortbl}
\graphicspath{{./figs/}} 

\usepackage{amsmath} 
\usepackage{amssymb}
\usepackage{amsfonts}
\usepackage{upgreek} 

\newcommand*\patchAmsMathEnvironmentForLineno[1]{%
\expandafter\let\csname old#1\expandafter\endcsname\csname #1\endcsname
\expandafter\let\csname oldend#1\expandafter\endcsname\csname
end#1\endcsname
 \renewenvironment{#1}%
   {\linenomath\csname old#1\endcsname}%
   {\csname oldend#1\endcsname\endlinenomath}%
}
\newcommand*\patchBothAmsMathEnvironmentsForLineno[1]{%
  \patchAmsMathEnvironmentForLineno{#1}%
  \patchAmsMathEnvironmentForLineno{#1*}%
}
\AtBeginDocument{%
\patchBothAmsMathEnvironmentsForLineno{equation}%
\patchBothAmsMathEnvironmentsForLineno{align}%
\patchBothAmsMathEnvironmentsForLineno{flalign}%
\patchBothAmsMathEnvironmentsForLineno{alignat}%
\patchBothAmsMathEnvironmentsForLineno{gather}%
\patchBothAmsMathEnvironmentsForLineno{multline}%
\patchBothAmsMathEnvironmentsForLineno{eqnarray}%
}

\usepackage[pdftex,
            pdfauthor={\paperauthors},
            pdftitle={\paperasciititle},
            pdfkeywords={\paperkeywords}]{hyperref}
\usepackage{hyperxmp}
\hypersetup{
    pdfcopyright={Copyright (C) \papercopyright},
    pdflicenseurl={\paperlicenceurl}
}

\usepackage[colorinlistoftodos,textsize=scriptsize]{todonotes}

\usepackage[bottom,flushmargin,hang,multiple]{footmisc}

\usepackage[all]{hypcap} 

\usepackage{xspace} 
\usepackage{upgreek}

\def\lhcb   {\mbox{LHCb}\xspace}

\def\belle  {\mbox{Belle}\xspace}

\def\MagUp {\mbox{\em Mag\kern -0.05em Up}\xspace}

\ifthenelse{\boolean{uprightparticles}}%
{

 \def\Ppi         {\ensuremath{\uppi}\xspace}

 \def\PDelta      {\ensuremath{\Delta}\xspace}                 
 \def\PXi         {\ensuremath{\Xi}\xspace}                 
 \def\PLambda     {\ensuremath{\Lambda}\xspace}                 
 \def\PSigma      {\ensuremath{\Sigma}\xspace}                 
 \def\POmega      {\ensuremath{\Omega}\xspace}                 
 \def\PUpsilon    {\ensuremath{\Upsilon}\xspace}
 \let\oldPi\Pi
 \def\PPi         {\ensuremath{\oldPi}\xspace}

 \def\PB      {\ensuremath{\mathrm{B}}\xspace}                 
 \def\PD      {\ensuremath{\mathrm{D}}\xspace}                 

 \def\PK      {\ensuremath{\mathrm{K}}\xspace}                 
 \def\Pb      {\ensuremath{\mathrm{b}}\xspace}                 
 \def\Pc      {\ensuremath{\mathrm{c}}\xspace}                 
 \def\Pd      {\ensuremath{\mathrm{d}}\xspace}                 

 \def\Pp      {\ensuremath{\mathrm{p}}\xspace}                 

 \def\Ps      {\ensuremath{\mathrm{s}}\xspace}

 \def\thebaroffset{0.0em}
}
{

 \def\Ppi         {\ensuremath{\pi}\xspace}

 \mathchardef\PDelta="7101
 \mathchardef\PXi="7104
 \mathchardef\PLambda="7103
 \mathchardef\PSigma="7106
 \mathchardef\POmega="710A
 \mathchardef\PUpsilon="7107
 \mathchardef\PPi="7105
 \def\PB      {\ensuremath{B}\xspace}                 
 \def\PD      {\ensuremath{D}\xspace}                 

 \def\PK      {\ensuremath{K}\xspace}                 
 \def\Pb      {\ensuremath{b}\xspace}                 
 \def\Pc      {\ensuremath{c}\xspace}                 
 \def\Pd      {\ensuremath{d}\xspace}                 

 \def\Pp      {\ensuremath{p}\xspace}                 

 \def\Ps      {\ensuremath{s}\xspace}

 \def\thebaroffset{0.18em}
}
\newcommand{\offsetoverline}[2][\thebaroffset]{\kern #1\overline{\kern -#1 #2}}%

\makeatletter
\ifcase \@ptsize \relax
  \newcommand{\miniscule}{\@setfontsize\miniscule{4}{5}}
\or
  \newcommand{\miniscule}{\@setfontsize\miniscule{5}{6}}
\or
  \newcommand{\miniscule}{\@setfontsize\miniscule{5}{6}}
\fi
\makeatother

\DeclareRobustCommand{\optbar}[1]{\shortstack{{\miniscule (\rule[.5ex]{1.25em}{.18mm})}
  \\ [-.7ex] $#1$}}

\def\dquark    {{\ensuremath{\Pd}}\xspace}

\def\squark    {{\ensuremath{\Ps}}\xspace}

\def\cquark    {{\ensuremath{\Pc}}\xspace}

\def\bquark    {{\ensuremath{\Pb}}\xspace}

\def\pion   {{\ensuremath{\Ppi}}\xspace}

\def\pip    {{\ensuremath{\pion^+}}\xspace}
\def\pim    {{\ensuremath{\pion^-}}\xspace}

\def\kaon    {{\ensuremath{\PK}}\xspace}

\def\KorKbar {\kern \thebaroffset\optbar{\kern -\thebaroffset \PK}{}\xspace}

\def\Kp      {{\ensuremath{\kaon^+}}\xspace}
\def\Km      {{\ensuremath{\kaon^-}}\xspace}

\def\D       {{\ensuremath{\PD}}\xspace}

\def\DorDbar {\kern \thebaroffset\optbar{\kern -\thebaroffset \PD}\xspace}

\def\Dp      {{\ensuremath{\D^+}}\xspace}
\def\Dm      {{\ensuremath{\D^-}}\xspace}

\def\DpDm    {\ensuremath{\Dp {\kern -0.16em \Dm}}\xspace}

\def\Ds      {{\ensuremath{\D^+_\squark}}\xspace}
\def\Dsp     {{\ensuremath{\D^+_\squark}}\xspace}
\def\Dsm     {{\ensuremath{\D^-_\squark}}\xspace}

\def\DporDsp {{\ensuremath{\D_{(\squark)}^+}}\xspace}

\def\B       {{\ensuremath{\PB}}\xspace}
\def\Bbar    {{\ensuremath{\offsetoverline{\PB}}}\xspace}

\def\BorBbar {\kern \thebaroffset\optbar{\kern -\thebaroffset \PB}\xspace}

\def\Bzb     {{\ensuremath{\Bbar{}^0}}\xspace}
\def\Bd      {{\ensuremath{\B^0}}\xspace}
\def\Bdb     {{\ensuremath{\Bbar{}^0}}\xspace}
\def\BdorBdbar {\kern \thebaroffset\optbar{\kern -\thebaroffset \Bd}\xspace}
\def\Bu      {{\ensuremath{\B^+}}\xspace}
\def\Bub     {{\ensuremath{\B^-}}\xspace}

\def\Bm      {{\ensuremath{\Bub}}\xspace}

\def\Bs      {{\ensuremath{\B^0_\squark}}\xspace}
\def\Bsb     {{\ensuremath{\Bbar{}^0_\squark}}\xspace}
\def\BsorBsbar {\kern \thebaroffset\optbar{\kern -\thebaroffset \Bs}\xspace}

\def\Bds     {{\ensuremath{\B_{(\squark)}^0}}\xspace}
\def\Bdsb    {{\ensuremath{\Bbar{}_{(\squark)}^0}}\xspace}

\def\Y#1S{\ensuremath{\PUpsilon{(#1S)}}\xspace}

\def\proton      {{\ensuremath{\Pp}}\xspace}
\def\antiproton  {{\ensuremath{\overline \proton}}\xspace}

\def\Lz          {{\ensuremath{\PLambda}}\xspace}
\def\Lbar        {{\ensuremath{\offsetoverline{\PLambda}}}\xspace}
\def\LorLbar     {\kern \thebaroffset\optbar{\kern -\thebaroffset \PLambda}\xspace}

\def\Xires       {{\ensuremath{\PXi}}\xspace}

\def\Xiresbar       {{\ensuremath{\offsetoverline{\Xires}}}\xspace}

\def\Lc          {{\ensuremath{\Lz^+_\cquark}}\xspace}
\def\Lcbar       {{\ensuremath{\Lbar{}^-_\cquark}}\xspace}

\def\Xicz        {{\ensuremath{\Xires^0_\cquark}}\xspace}
\def\Xicp        {{\ensuremath{\Xires^+_\cquark}}\xspace}

\def\Xicbarm     {{\ensuremath{\Xiresbar{}_\cquark^-}}\xspace}

\def\BF         {{\ensuremath{\mathcal{B}}}\xspace}

\newcommand{\decay}[2]{\mbox{\ensuremath{#1\!\to #2}}\xspace} 

\def\to                 {\ensuremath{\rightarrow}\xspace}

\def\grpsuthree {{\ensuremath{\mathrm{SU}(3)}}\xspace}

\def\CP                {{\ensuremath{C\!P}}\xspace}

\def\Vcd  {{\ensuremath{V_{\cquark\dquark}^{\phantom{\ast}}}}\xspace}

\def\Vcs  {{\ensuremath{V_{\cquark\squark}^{\phantom{\ast}}}}\xspace}

\def\AT#1     {\ensuremath{A_{\mathrm{T}}^{#1}}\xspace}           

\def\C#1      {\ensuremath{\mathcal{C}_{#1}}\xspace}                       
\def\Cp#1     {\ensuremath{\mathcal{C}_{#1}^{'}}\xspace}                    
\def\Ceff#1   {\ensuremath{\mathcal{C}_{#1}^{\mathrm{(eff)}}}\xspace}        
\def\Cpeff#1  {\ensuremath{\mathcal{C}_{#1}^{'\mathrm{(eff)}}}\xspace}       
\def\Ope#1    {\ensuremath{\mathcal{O}_{#1}}\xspace}                       
\def\Opep#1   {\ensuremath{\mathcal{O}_{#1}^{'}}\xspace}                    

\newcommand{\aunit}[1]{\ensuremath{\text{\,#1}}}       

\newcommand{\tev}{\aunit{Te\kern -0.1em V}\xspace}
\newcommand{\gev}{\aunit{Ge\kern -0.1em V}\xspace}
\newcommand{\mev}{\aunit{Me\kern -0.1em V}\xspace}
\newcommand{\kev}{\aunit{ke\kern -0.1em V}\xspace}
\newcommand{\ev}{\aunit{e\kern -0.1em V}\xspace}
 
\newcommand{\mevc}{\ensuremath{\aunit{Me\kern -0.1em V\!/}c}\xspace}
\newcommand{\gevc}{\ensuremath{\aunit{Ge\kern -0.1em V\!/}c}\xspace}
\newcommand{\mevcc}{\ensuremath{\aunit{Me\kern -0.1em V\!/}c^2}\xspace}
\newcommand{\gevcc}{\ensuremath{\aunit{Ge\kern -0.1em V\!/}c^2}\xspace}

\def\fb   {\ensuremath{\aunit{fb}}\xspace}
\def\invfb   {\ensuremath{\fb^{-1}}\xspace}

\newcommand{\chisq}{\ensuremath{\chi^2}\xspace}

\newcommand{\chisqip}{\ensuremath{\chi^2_{\text{IP}}}\xspace}

\def\gsim{{~\raise.15em\hbox{$>$}\kern-.85em
          \lower.35em\hbox{$\sim$}~}\xspace}
\def\lsim{{~\raise.15em\hbox{$<$}\kern-.85em
          \lower.35em\hbox{$\sim$}~}\xspace}

\def\sPlot{\mbox{\em sPlot}\xspace}

\def\pt         {\ensuremath{p_{\mathrm{T}}}\xspace}

\def\mrad{\aunit{mrad}\xspace}

\def\evtgen     {\mbox{\textsc{EvtGen}}\xspace}

\def\geant      {\mbox{\textsc{Geant4}}\xspace}

\def\photos     {\mbox{\textsc{Photos}}\xspace}

\def\pythia     {\mbox{\textsc{Pythia}}\xspace}

\def\cpp        {\mbox{\textsc{C\raisebox{0.1em}{{\footnotesize{++}}}}}\xspace}

\def\tell1  {TELL1\xspace}
\def\ukl1   {UKL1\xspace}

\newcommand{\lhcborcid}[1]{\href{https://orcid.org/#1}{\hspace*{0.1em}\raisebox{-0.45ex}{\includegraphics[width=1em]{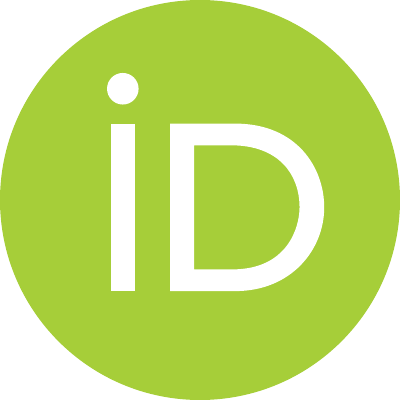}}}}

\hypersetup{
  colorlinks   = true, 
  urlcolor     = blue, 
  linkcolor    = blue, 
  citecolor    = red   
}

\ifEnableSectionTOCLinks
    \usepackage[explicit]{titlesec} 
    
    \let\oldcontentsline\contentsline
    \renewcommand

    \titleformat{\section}{\normalfont\Large\bf}{\hyperlink{tocsection.\thesection}{{\thesection} \parbox[t]{\dimexpr\textwidth-1pc}{#1}}}{1pc}{}

    \titleformat{\subsection}{\normalfont\bf}{\hyperlink{tocsubsection.\thesubsection}{{\thesubsection} \parbox[t]{\dimexpr\textwidth-1pc}{#1}}}{1pc}{}

    \titleformat{name=\section,numberless}[display]{}{}{0pt}{\normalfont\Huge\bfseries #1}
\fi

\usepackage{cite} 
\usepackage{mciteplus}

\newcommand{\LcLcbar}{\Lc \Lcbar}
\def\Bds{{\ensuremath{\B^0_{\kern -0.1em{\scriptscriptstyle (}\kern -0.05em\squark\kern -0.03em{\scriptscriptstyle )}}}}\xspace}
\def\Bdsbar{{\ensuremath{\Bbar{}^0_{\kern -0.1em{\scriptscriptstyle (}\kern -0.05em\squark\kern -0.03em{\scriptscriptstyle )}}}}\xspace}

\newcommand{\BdDsD}{\decay{\Bdb}{\Dsm \Dp}}

\newcommand{\BdDsDKKpiKpipi}{\decay{\Bdb}{\Dsm(\Kp\Km\pim) \Dp(\Km\pip\pip)}}

\newcommand{\BdLcLcbar}{\decay{\Bdb}{\Lc \Lcbar}}

\newcommand{\BLcLcbar}{\decay{\Bdsbar}{\Lc \Lcbar}}

\newcommand{\BsDsD}{\decay{\Bs}{\Dsm \Dp}}
\newcommand{\BsDsDs}{\decay{\Bsb}{\Dsp \Dsm}}

\newcommand{\BsDsDsKKpiKKpi}{\decay{\Bsb}{\Dsp(\Km\Kp\pip) \Dsm(\Kp\Km\pim)}}

\newcommand{\BsLcLcbar}{\decay{\Bsb}{\Lc \Lcbar}}

\newcommand{\DsKKpi}{\decay{\Dsp}{\Km\Kp\pip}}
\newcommand{\DKpipi}{\decay{\Dp}{\Km\pip\pip}}
\newcommand{\LcpKpi}{\decay{\Lc}{p\Km\pip}}

\newcommand{\BsPPbar}{\decay{\Bsb}{\proton \antiproton}}

\newcommand{\zhushi}[1]{}
\newcommand{\BdXicXicbar}{\decay{\Bdb}{\Xicp \Xicbarm}}

\newcommand{\BdXicLcbar}{\decay{\Bdb}{\Xicp \Lcbar}}

\def\prodprobnnx     {\ensuremath{\prod_i{\mathcal{P}_i}}\xspace}

\def\FOMp    {\ensuremath{\epsilon/(\frac{\sigma}{2}+\sqrt{N_\text{B}})}\xspace}

\usepackage{caption,subcaption}
\begin{document}

\renewcommand{\thefootnote}{\fnsymbol{footnote}}
\setcounter{footnote}{1}


\begin{titlepage}
\pagenumbering{roman}

\vspace*{-1.5cm}
\centerline{\large EUROPEAN ORGANIZATION FOR NUCLEAR RESEARCH (CERN)}
\vspace*{1.5cm}
\noindent
\begin{tabular*}{\linewidth}{lc@{\extracolsep{\fill}}r@{\extracolsep{0pt}}}
\vspace*{-1.5cm}\mbox{\!\!\!\includegraphics[width=.14\textwidth]{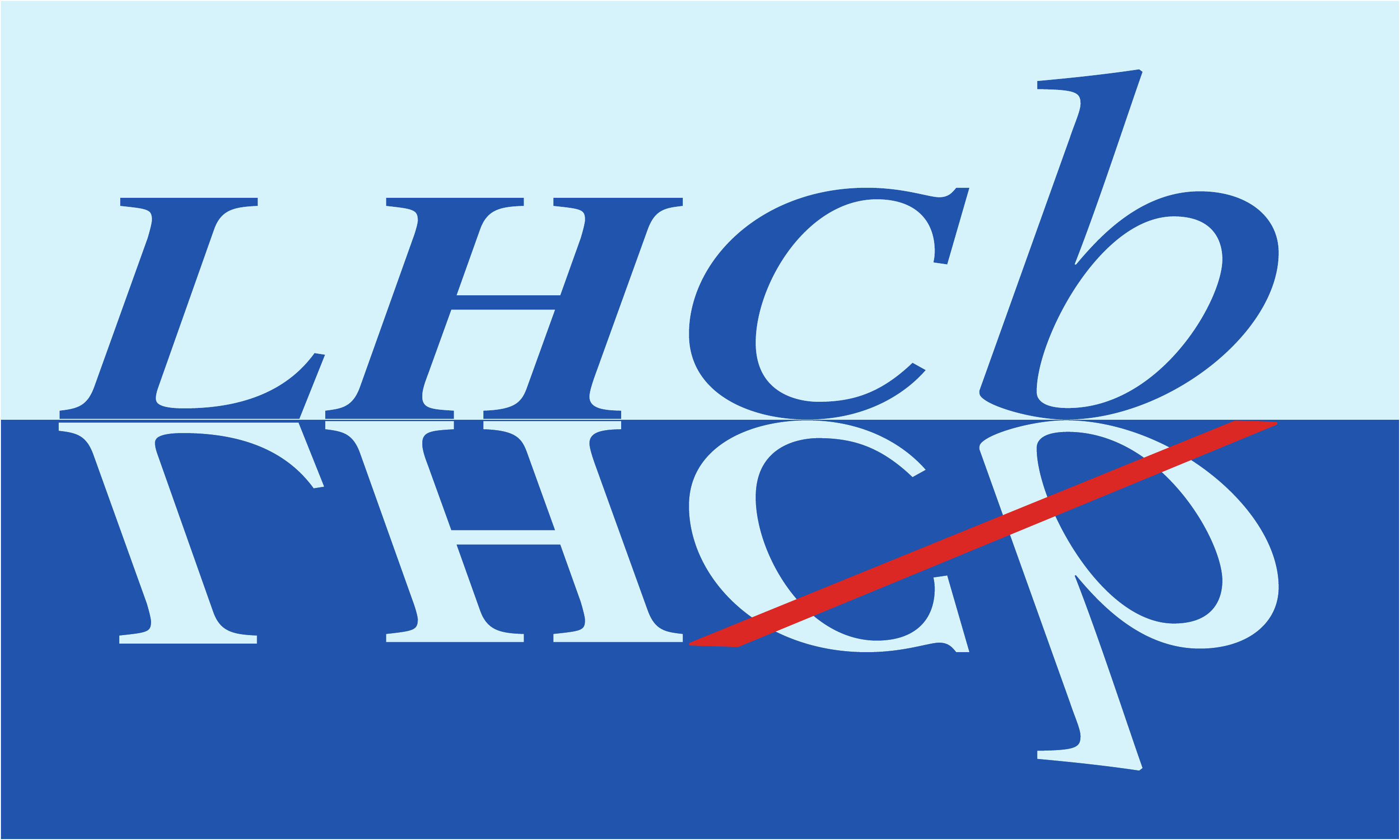}} & &
\\
 & & CERN-EP-2025-258 \\  
 & & LHCb-PAPER-2025-053 \\  
 & & February 13, 2026 \\ 
 & & \\
\end{tabular*}

\vspace*{4.0cm}

{\normalfont\bfseries\boldmath\huge
\begin{center}
  \papertitle 
\end{center}
}

\vspace*{1.0cm}

\begin{center}
\paperauthors\footnote{Authors are listed at the end of this Letter.}
\end{center}
\vspace{\fill}
\begin{abstract}
  \noindent
  A search is presented for the two-body charmed baryonic decays, $\Bdsb\to\Lc\Lcbar$, using a data sample collected by the LHCb experiment during 2011--2012 and 2015--2018, corresponding to an integrated luminosity of $9\invfb$. The first observation of the $\Bsb\to\Lc\Lcbar$ decay is reported with $6.2\sigma$ significance along with $4.3\sigma$ evidence for the $\Bdb\to\Lc\Lcbar$ decay. The branching fractions are measured to be $\BF(\Bdb\to\Lc\Lcbar) = (1.01^{+0.27}_{-0.28} \pm 0.08 \pm 0.15) \times 10^{-5}$ and $\BF(\Bsb\to\Lc\Lcbar) = (5.0 \pm 1.3 \pm 0.5 \pm 0.8) \times 10^{-5}$, where the first uncertainty is statistical, the second systematic, and the third due to external inputs. These results provide novel experimental inputs for the theoretical framework describing two-body baryonic decays of $B$ mesons via $W$-emission and $W$-exchange mechanisms.
\end{abstract}

\vspace*{2.0cm}

\begin{center}
  Published in Phys.~Rev.~Lett. 136 (2026) 061802
\end{center}

\vspace{\fill}

{\footnotesize 
\centerline{\copyright~\papercopyright. \href{\paperlicenceurl}{\paperlicence}.}}
\vspace*{2mm}

\end{titlepage}


\newpage
\setcounter{page}{2}
\mbox{~}
%
%
%
%


\renewcommand{\thefootnote}{\arabic{footnote}}
\setcounter{footnote}{0}


\cleardoublepage


\pagestyle{plain} 
\setcounter{page}{1}
\pagenumbering{arabic}


The study of \B-meson decays has long been a fertile ground for testing the standard model of particle physics and probing for potential new physics beyond it. 
Among the diverse decay channels, baryonic decays of \B mesons provide a unique laboratory for exploring the interplay between weak interactions and quantum chromodynamics in the nonperturbative regime~\cite{Cheng:2006nm,Huang:2021qld}. 
These decays are not only sensitive to the Cabibbo-Kobayashi-Maskawa matrix elements~\cite{Cabibbo:1963yz,Kobayashi:1973fv}, but they also offer insights into the dynamics of baryon production and final-state interactions, which are less understood compared to mesonic systems.

\begin{figure}[!bp]
    \centering
   \subfloat{\includegraphics[width=0.47\textwidth]{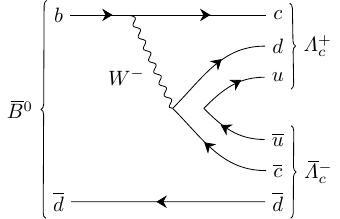}}
      \subfloat{\includegraphics[width=0.5\textwidth]{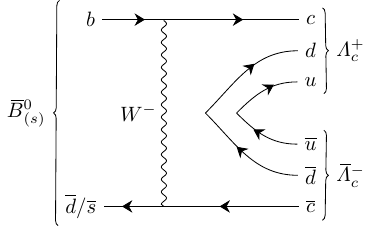}}
    \caption{Tree-level Feynman diagrams for $\Bdsb\to\Lc\Lcbar$ decays. The left diagram shows the contribution from the $W$-emission process, while the right one depicts the $W$-exchange process.}
    \label{fig:Feynmann}
\end{figure}
In the standard model, two-body baryonic \B-meson decays proceed at tree level through the nonfactorizable internal $W$-emission quark diagram and the factorizable $W$-exchange and $W$-annihilation topology.
However, the $W$-exchange and -annihilation processes can be helicity suppressed~\cite{Chen:2008pf,BaBar:2014omp}, and hence, are often neglected in theoretical studies~\cite{Chernyak:1990ag,Cheng:2001tr,Chang:2001jt,Chua:2013zga,He:2006vz,Cheng:2009yz,Geng:2025yna}. Experimental searches for the factorizable topology are currently limited to the \BsPPbar decay,\footnote{Charge-conjugate decays are implied throughout.} with an upper limit on the branching fraction \BF set at \mbox{$4.4 \times 10^{-9}$} at $90\%$ confidence level (CL)~\cite{LHCb-PAPER-2022-004}.
However, other theories suggest that processes with quark-level \mbox{$\bquark\to\cquark$} transitions, such as \BsLcLcbar, \BdXicXicbar, and \decay{\Bsb}{\proton\Lcbar}, can circumvent severe helicity suppression, making the factorizable topology non-negligible~\cite{Hsiao:2019wyd,Hsiao:2023mud,Geng:2024uxp} and potentially leading to sizable direct \CP asymmetries~\cite{Geng:2025yna}.

As illustrated in the Feynman diagrams in Fig.~\ref{fig:Feynmann}, the two-body charmed decays \BdLcLcbar are expected to proceed mainly through tree-level $W$-emission amplitudes, while \BsLcLcbar decays are expected to proceed via suppressed topologies~\cite{Rui:2024xgc}.
As a consequence, the \BdLcLcbar branching fraction is expected to be larger than that of the corresponding \Bsb-meson decay. If the \BdLcLcbar decay proceeds similarly to \decay{\Bm}{\Xicz\Lcbar} and \BdXicLcbar decays, which are also believed to be dominated by the nonfactorizable $W$-emission topology, then its branching fraction is expected to be \mbox{$|\Vcd/\Vcs|^2 (\tau_{\Bd}/\tau_{\Bu}) \BF(\decay{\Bm}{\Xicz\Lcbar}) = (4.7 \pm 1.1) \times 10^{-5}$} in the
 $U$-spin symmetry limit. Here, the CKM matrix elements \Vcd and \Vcs, along with the \Bd and \Bu meson lifetimes $\tau_{\Bd}$ and $\tau_{\Bu}$, are taken from Ref.~\cite{PDG2024}. While the \belle collaboration has reported an upper limit on the \BdLcLcbar branching fraction of \mbox{$6.2 \times 10^{-5}$} at 90\% CL~\cite{Belle:2007lyc}, consistent with this prediction, the LHCb collaboration has determined more strict upper limits of \mbox{$1.6 \times 10^{-5}$} and \mbox{$8.0 \times 10^{-5}$} at $95\%$ CL for the \BdLcLcbar and \BsLcLcbar branching fractions, respectively~\cite{LHCb-PAPER-2014-002}. These measured upper limits suggest a tension between theory and experiment, indicating a contribution from $W$-exchange or interference effects, which requires more precision measurements for confirmation. 

This Letter presents a search for \mbox{\decay{\Bdsb}{\Lc(\proton \Km \pip)\Lcbar(\antiproton \Kp \pim)}} decays using proton-proton ($pp$) collision data collected with the \lhcb detector at center-of-mass energies of 7, 8, and 13\tev during the LHC Run~1 (2011--2012) and Run~2 (2015--2018) data-taking periods, corresponding to a total integrated luminosity of $9\invfb$. The branching fractions of the signal decays are measured relative to the topologically similar \mbox{\BdDsDKKpiKpipi} and \mbox{\BsDsDsKKpiKKpi} normalization channels using
\begin{eqnarray*}
\label{eq:BF_vis_BdLcLc}
\BF(\BdLcLcbar)&=&\BF(\BdDsD)\cdot  \frac{\BF_{\Dsp}\BF_{\Dp}}{\BF^2_{\Lc}} \cdot \frac{N(\BdLcLcbar)}{N(\BdDsD)}\cdot\frac{\epsilon(\BdDsD)}{\epsilon(\BdLcLcbar)},\\
\label{eq:BF_vis_BsLcLc}
\BF(\BsLcLcbar)&=&\BF(\BsDsDs)\cdot  \frac{\BF^2_{\Dsp}}{\BF^2_{\Lc}} \cdot \frac{N(\BsLcLcbar)}{N(\BsDsDs)} \cdot \frac{\epsilon(\BsDsDs)}{\epsilon(\BsLcLcbar)},
\end{eqnarray*}
where $N$ denotes the measured yield; $\epsilon$ represents the combined efficiency, including the detector acceptance, trigger, reconstruction, and selection efficiencies; and $\mathcal{B}_{\Dsp}$, $\mathcal{B}_{\Dp}$, and $\mathcal{B}_{\Lc}$ denote the known branching fractions of \DsKKpi, \DKpipi, and \LcpKpi decays, respectively~\cite{PDG2024}.

The LHCb detector is a 
single-arm forward spectrometer covering the pseudorapidity range 
$2 < \eta < 5$, 
described in detail in Refs.~\cite{LHCb-DP-2008-001,LHCb-DP-2014-002}. The online event selection is performed by a trigger~\cite{LHCb-DP-2012-004,LHCb-DP-2019-001}, which consists of a hardware stage followed by a two-level software stage that applies a full event reconstruction. At the hardware trigger stage, events are required to have a hadron, photon, or electron with high transverse energy in the calorimeters. For hadrons, the transverse energy threshold is 3.5\gev. At the software trigger stage, there must be a secondary vertex consistent with a two-, three-, or four-body decay topology, with significant displacement from any primary $pp$ collision vertex (PV). At least one charged particle must have a transverse momentum $\pt > 1.6\gev$ and be inconsistent with originating from any PV.\footnote{Natural units with $\hbar=c=1$ are used throughout this Letter.} 

Simulated samples are used to study the properties of the signal, normalization, and background channels. Proton-proton collisions are generated by \pythia~\cite{Sjostrand:2007gs} with a specific \lhcb configuration~\cite{LHCb-PROC-2010-056}.
Decays of unstable particles are described by \evtgen~\cite{Lange:2001uf}, in which final-state radiation is generated using \photos~\cite{Photos-2}. The interactions of the generated particles with the detector material, and their responses, are implemented using the \geant toolkit~\cite{Agostinelli:2002hh,Allison:2006ve}, as described in Ref.~\cite{LHCb-PROC-2011-006}. 

In the offline selection, the candidate decays are reconstructed by first forming \Lc (or $\DporDsp$) candidates from three-track combinations and then combining pairs of these charmed hadrons into \Bdsb mesons. Each final-state particle must fulfill the momentum requirements $p>5\gev$, $\pt>500\mev$, and must be inconsistent with originating from any PV using the criterion $\chisqip > 4$, where \chisqip is the difference between the vertex-fit \chisq of a given PV reconstructed with and without the particle in question. For proton candidates, a more stringent requirement $p>10\gev$ is imposed. In addition, a loose particle identification (PID) requirement is applied to all charged tracks~\cite{LHCb-DP-2012-003,LHCb-PUB-2016-021}. The charmed hadrons are required to satisfy $\pt>1\gev$ with a good-quality decay vertex significantly separated from any PV. To further suppress singly charmed and charmless background sources, the measured decay length along the beamline direction for each charmed hadron candidate must be greater than half of its resolution. Subsequently, a good \Bdsb decay vertex is required, which must be significantly displaced from the associated PV, defined as the PV for which the particle has the smallest \chisqip. The momentum of the \Bdsb candidate is required to be consistent with its flight direction.
To suppress background arising from the misreconstruction of a single charged-particle trajectory as two or more distinct tracks, candidate decays are rejected if the opening angle between the momentum vectors of any pair of final-state tracks is less than 0.5\mrad. 

To suppress cross-feed among $\Lc$, $\Ds$, and $\Dp$ candidates, veto criteria are implemented based on invariant-mass and PID requirements. 
For \mbox{\LcpKpi} decays, candidates are rejected if the \mbox{$\proton\Km$} invariant mass, after assigning the kaon mass to the proton candidate, lies within $\pm10\mev$ of the known $\phi(1020)$ mass~\cite{PDG2024}. Additional PID requirements are applied when the reconstructed \mbox{$\proton\Km\pip$} invariant mass lies within $\pm25\mev$ of the known \Dp (\Ds) mass after assigning the pion (kaon) mass hypothesis to the proton candidate.
  Similarly, a tighter PID requirement is applied to the \Kp meson in \mbox{\DsKKpi} decays for events where the invariant mass of the \mbox{\Km\Kp\pip} combination, computed after assigning the pion (proton) mass to the kaon, falls within $\pm25\mev$ of the \Dp (\Lc) mass. In the case of \mbox{\DKpipi} decays, after assigning the mass of either of the two pions to the kaon (proton), candidates with \mbox{$\Km\pip\pip$} invariant mass within $\pm25\mev$ of the $\Ds$ (\Lc) mass are rejected unless a tighter PID requirement is fulfilled.

The two most powerful variables for discriminating between signal and the remaining combinatorial background are the $\chisqip$ of the \Bdsb candidate with respect to its PV and the quantity \prodprobnnx. The latter is defined as the product of the neural-network-based PID variables for correct identification of all six final-state particles~\cite{LHCb-DP-2014-002}. As the PID algorithms are tuned individually for Run~1 and Run~2 operations, the \prodprobnnx optimization is performed separately for each data-taking period. The selection criteria for these two variables are determined using data from the sideband region, \mbox{$5150 < m(\LcLcbar) < 5240\mev$} or \mbox{$5407 < m(\LcLcbar) <5550\mev$} as background proxy, and simulated \BsLcLcbar samples as signal proxy, with the goal of maximizing the Punzi figure of merit $\FOMp$~\cite{Punzi:2003bu}. Here, $\epsilon$ represents the signal efficiency for a given selection criterion, $\sigma$ denotes the target signal significance (set to $5$), and $N_\text{B}$ corresponds to the background yield in the signal region \mbox{$5350 < m(\LcLcbar) <5380\mev$}. The optimized $\prodprobnnx$ and $\chisqip$ selection criteria, determined separately for Run~1 and Run~2, are applied to all data and simulation samples.

To ensure a realistic description of the data, the kinematic distributions and the PID variables of charged tracks in simulated samples are corrected. The kinematic correction accounts for dependencies on $\pt$, $\eta$, event multiplicity, and $\chisqip$ of the \Bdsb candidate. The correction weights are determined through a data-driven approach employing a gradient-boosting algorithm~\cite{Rogozhnikov:2016bdp}, where background-subtracted data samples are obtained using the \sPlot technique~\cite{Pivk:2004ty} applied to normalization decay modes. 
The PID variables are calibrated via a kinematic-dependent transformation, where a kernel density estimation (KDE) algorithm is used to provide a smooth interpolation of distributions from dedicated calibration samples~\cite{LHCb-PUB-2016-021}.

Efficiencies are derived from simulated samples after applying the corrections outlined above and imposing all selection criteria. The efficiency ratios between signal and normalization channels are determined separately for each data-taking year, and then averaged according to the integrated luminosity of each sample, giving overall values  \mbox{$\epsilon(\BdLcLcbar)/{\epsilon(\BdDsD)}=0.67\pm0.01$} and \mbox{$\epsilon(\BsLcLcbar)/{\epsilon(\BsDsDs)}=0.56\pm0.01$}. The uncertainties are purely due to the limited size of the simulation samples.
\begin{figure}[tb]
    \centering
    \includegraphics[width=0.6\textwidth]{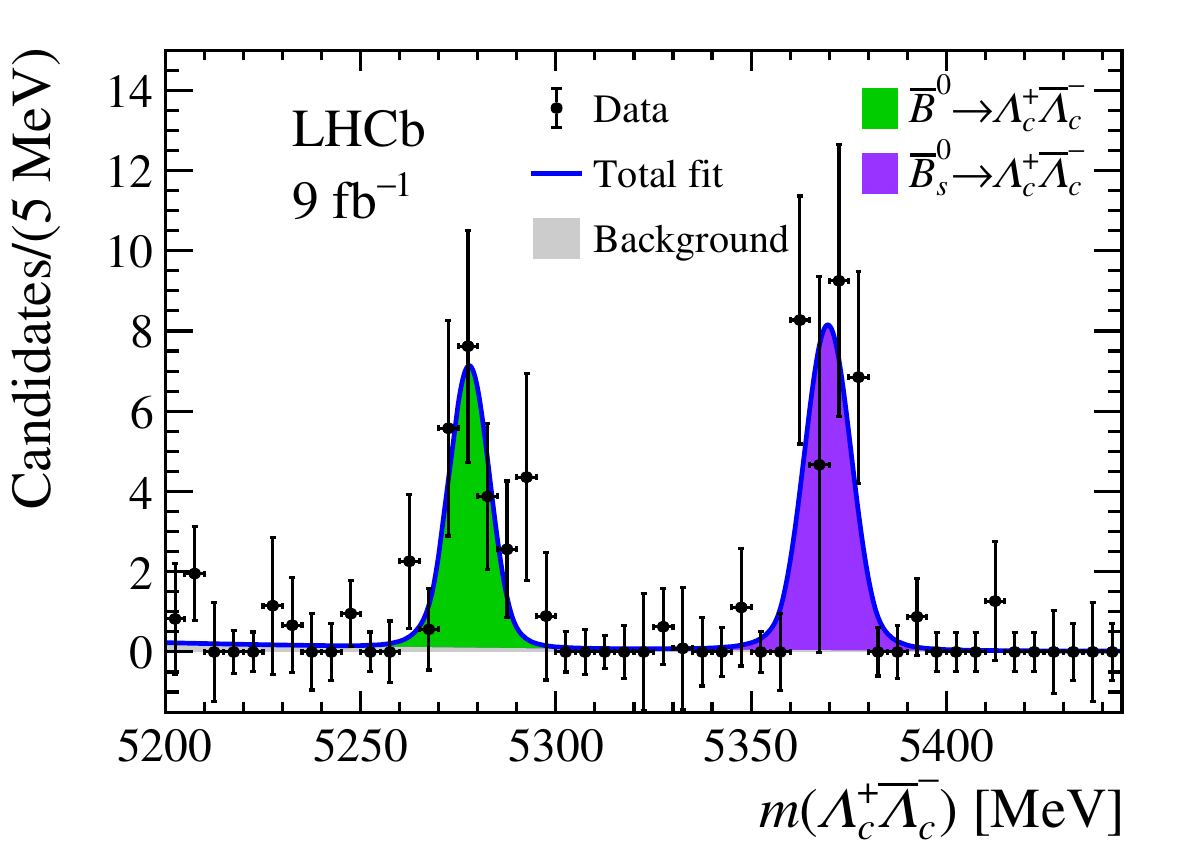}
    \caption{Mass distribution of the $\Lc\Lcbar$ system with singly charmed and charmless backgrounds subtracted. The fit result (blue curve) is shown together with the signal components from $\Bdb\to\Lc\Lcbar$ (filled green) and $\Bsb\to\Lc\Lcbar$ (filled purple) decays, and the nonresonant background contribution (filled gray).}    
    \label{fig:SignalMasses}
\end{figure}

The \BLcLcbar yields are obtained from a two-stage fitting procedure. The first stage is an unbinned maximum-likelihood fit to the two-dimensional \mbox{$[m(\Lc),m(\Lcbar)]$} distribution, aiming to separate genuine \LcLcbar candidates from charmless and singly charmed backgrounds. This fit is performed in 5\mev bins of $m(\Lc\Lcbar)$ to determine background-subtracted yields. 
In this fit, the \Lc or \Lcbar signal is described by a modified crystal ball distribution~\cite{Skwarnicki:1986xj} with tails on both sides of the Gaussian core (double-sided crystal ball, DSCB), where the tail parameters are fixed to those found in simulation, and the peak position is allowed to vary among different $m(\Lc\Lcbar)$ bins to account for possible correlations between the $\Bdsb$ mass and the \Lc(\Lcbar) mass. In the baseline fit, the ratio of widths of the \Lc peak between data and simulation is constrained with Gaussian priors to the value derived from the normalization channel. The background in each dimension is described by an exponential function. The construction of the fit model and the corresponding results are detailed in the End Matter.

 The second stage is a binned maximum-likelihood fit to the resulting $m(\Lc\Lcbar)$ distribution of the doubly charmed decays, as shown in Fig.~\ref{fig:SignalMasses}, to determine the \BdLcLcbar and \BsLcLcbar signal yields, with potential background from nonpeaking \LcLcbar also included.  The $\Bdsb$ signal shape is described by a DSCB distribution with the tail parameters fixed from simulation. Widths of the $\Bdb$ and $\Bsb$ peaks are fixed to the values measured for the corresponding normalization channels scaled by the ratio of signal to normalization widths determined from simulation. The background shape is assumed to be exponential, with the slope parameter free to vary in the fit. The \BdLcLcbar and \BsLcLcbar signal yields obtained from the fit to the samples are \mbox{$19.0^{+5.0}_{-5.2}$} and \mbox{$25.1 \pm 6.7$}, respectively, where the uncertainties are statistical only and determined from pseudoexperiments.

The normalization channel yields are obtained using a similar two-stage fitting procedure as for the signal channels, where the $\Bdsb$ and $\DporDsp$ signal and background components are modeled using the same line shapes as the \BLcLcbar decays. The line shape for the misidentified \BdDsD decay in the \BsDsDs dataset is estimated from simulated samples using the KDE approach~\cite{Cranmer:2000du}. The fits yield \mbox{$26\,171\pm 164$} \BdDsD signal decays and \mbox{$2926\pm65$} \BsDsDs signal decays, with the fit projections provided in the End Matter.

Several sources of systematic uncertainty from the mass fit and the efficiency ratio are considered. A summary is presented in Table~\ref{tab:summary_syst_BF}.
Alternative mass fits are performed with different configurations of the fit model, and the largest change in the branching ratio is assigned as the corresponding systematic uncertainty. The signal models are varied to use either fixed or Gaussian-constrained widths for the \Bds peaks and either Gaussian-constrained or floated widths for the \Lc peak. Alternative signal models for \Lc, $D_{(\squark)}^+$, and \Bds peaks are tested either by varying the DSCB tail parameters or by replacing the DSCB distribution with a Hypatia~\cite{Santos:2013gra} distribution. The exponential background shapes are replaced with Chebyshev polynomials. The line shape of the misidentified \BdDsD decay in the \BsDsDs dataset is replaced by an alternative KDE with a 50\% wider sampling width. The fitting procedure is validated using pseudoexperiments. A bias at a level of a few percent is observed in the fitted signal yield, mainly arising from the positivity constraint on the signal yields in the first-stage fit and the inherent bias of maximum-likelihood fits with small yields~\cite{rice2006mat}. The size of this bias is assigned as a systematic uncertainty.  

\begin{table}[btp]
\caption{Relative systematic uncertainties on the measured branching fraction ratios. The dashes indicate sources that are not relevant for that measurement.}
\label{tab:summary_syst_BF}
\begin{center}
\begin{tabular}{ccc}
\hline
Source   & $\frac{\BF(\Bdb\to\Lc\Lcbar)}{\BF(\Bdb\to\Dsm\Dp)}$ & $\frac{\BF(\Bsb\to\Lc\Lcbar)}{\BF(\Bsb\to\Dsp\Dsm)}$ \\ 
\hline
    \Lc peak width  &  2.1\%  & 4.3\% \\
    $\Bdsb$ peak width & 0.2\% & 0.5\%   \\ 
      \Lc or $D_{(\squark)}^+$ line shape &  0.2\% & 0.2\% \\
 \Bdsb signal line shape &                      $3.6\%$ &                      $0.5\%$\\
      Background line shape &                      $1.4\%$ &                      $7.1\%$\\
            Line shape of misidentified decay & not applicable & 0.1\% \\
              Fit bias &                      $5.7\%$ &                      $1.2\%$\\
Tracking efficiency  &                 $1.3\%$ &                 $1.6\%$\\
PID correction &                 $0.7\%$ &                 $0.9\%$\\
 Simulated sample size&                 $1.7\%$ &                 $1.8\%$\\
Kinematical correction&                 $0.2\%$ &                 $0.3\%$\\
\Lc Dalitz plot and polarization correction & $3.6\%$ &  $4.3\%$ \\ 
\Bsb lifetime correction & not applicable & $0.6\%$ \\
         \hline
Total & 8.4\%  &  9.9\%  \\ 
\hline
\end{tabular} 
\end{center} 
\end{table}

Systematic uncertainties associated with the efficiencies estimated from simulation are also evaluated. Limited knowledge of the detector material budget and particle interaction cross sections leads to uncertainties on the tracking efficiencies, which are estimated using the approaches provided in Ref.~\cite{LHCb-DP-2013-002}. 
 The corrections applied to the PID variables in the simulation depend on both the sample size of the calibration datasets and the kernel densities used in the sampling. To evaluate the systematic uncertainty, the KDE template is varied by increasing the kernel width by $50\%$ and by implementing a bootstrapping technique. The uncertainties arising from the limited size of the simulated samples are estimated under the assumption that the efficiencies follow binomial distributions. A systematic uncertainty associated with the kinematical correction procedure is evaluated by performing an alternative correction that employs a different set of kinematic variables to train the weighting algorithm. The systematic uncertainty associated with imperfect simulation of the hardware trigger is estimated using a data-driven method~\cite{LHCb-DP-2012-004,LHCb-DP-2019-001} and found to be negligible. This is expected, given the similar kinematics and decay topologies between the signal and normalization decays. 
 
 To address uncertainties due to the treatment in simulation of intermediate states in the \decay{\Lc}{\proton\Km\pip} decay and the \Lc-baryon polarization, two weighting procedures are implemented. The \Lc Dalitz plot distributions in simulation are weighted to match those in data. Then, the \Lc polarization is emulated by a weight \mbox{$1 + \alpha P \cos\theta_p$}, where $\theta_p$ is the angle between the proton momentum in the \Lc rest frame and the \Lc direction in the $\Bdsb$ rest frame. The decay parameter $\alpha$ and polarization $P$ are taken from Ref.~\cite{LHCb-PAPER-2022-002}. The systematic uncertainty is evaluated as the relative difference between results with and without these weights applied. 
 An additional systematic uncertainty is assigned to account for the difference between the average \Bsb lifetime used in simulation and the physical lifetimes of the decay modes, which are dominated by the shorter-lived \Bsb mass eigenstate for both signal and normalization modes. 
 The effect is evaluated by reweighting the \Bsb lifetime distribution in simulated samples and found to be small.

After evaluating the systematic uncertainties, the branching fractions are measured as
\begin{align*}
\BF(\BdLcLcbar) &= (1.01^{+0.27}_{-0.28} \pm 0.08 \pm 0.15) \times 10^{-5}, \\
\BF(\BsLcLcbar) &= (5.0 \pm 1.3 \pm 0.5 \pm 0.8) \times 10^{-5},
\end{align*}
where the uncertainties, respectively, represent statistical, systematic, and external contributions from branching fractions of intermediate charmed decays and normalization channels~\cite{PDG2024}.

To evaluate the significance of each signal decay, a test statistic \mbox{$q = -2\ln(\mathcal{L}_0 / \hat{\mathcal{L}})$} is defined, where $\mathcal{L}_0$ represents the likelihood of the model with the contribution from either \BdLcLcbar or \BsLcLcbar removed, and $\hat{\mathcal{L}}$ corresponds to the likelihood when both contributions are included. Among all the fit variants discussed above, the model yielding the minimal $q$ value in the data is identified. Datasets from pseudoexperiments are then generated by excluding either \Bdb or \Bsb contributions from this model. The $p$ values associated with the observed \Bdb and \Bsb signals in the data are calculated from the tails of the two resulting $q$ distributions derived from these datasets and then converted to significances. The resulting significance, considering the systematic uncertainty, is $4.3\sigma$ for the \BdLcLcbar decay and $6.2\sigma$ for the \BsLcLcbar decay.

In summary, we present the first measurements of \BdLcLcbar and \BsLcLcbar branching fractions using the complete \lhcb Run~1 and Run~2 datasets. The first observation of the \BsLcLcbar decay represents the first experimental verification of the $W$-exchange process in baryonic $B$-meson decays, a mechanism long considered helicity suppressed, and therefore, often neglected. The measured value of $\BF(\BdLcLcbar)$ is at the level of \mbox{$1 \times 10^{-5}$}, indicating a significant discrepancy with the expected contribution from the $W$-emission diagram alone, as presented earlier in this Letter.
This deviation provides further evidence for the presence of \grpsuthree-breaking effects and a $W$-exchange contribution that interferes destructively with the internal $W$-emission process, thereby suppressing the observed branching fraction. 
Taken together, these observations suggest that the long-overlooked contribution from $W$-annihilation and -exchange mechanisms in charmless, singly charmed, and doubly charmed two-body baryonic $B$-meson decays needs a thorough reexamination. Such effects may be important for improving predictions of \CP-violating phenomena in baryonic $B$ decays, which could potentially be studied using data from LHCb Run~3 and future upgrades~\cite{LHCb-TDR-012,LHCb-DP-2022-002}.


\section*{Acknowledgements}
%
%
\noindent We express our gratitude to our colleagues in the CERN
accelerator departments for the excellent performance of the LHC. We
thank the technical and administrative staff at the LHCb
institutes.
We acknowledge support from CERN and from the national agencies:
ARC (Australia);
CAPES, CNPq, FAPERJ and FINEP (Brazil); 
MOST and NSFC (China); 
CNRS/IN2P3 (France); 
BMFTR, DFG and MPG (Germany);
INFN (Italy); 
NWO (Netherlands); 
MNiSW and NCN (Poland); 
MCID/IFA (Romania); 
MICIU and AEI (Spain);
SNSF and SER (Switzerland); 
NASU (Ukraine); 
STFC (United Kingdom); 
DOE NP and NSF (USA).
We acknowledge the computing resources that are provided by ARDC (Australia), 
CBPF (Brazil),
CERN, 
IHEP and LZU (China),
IN2P3 (France), 
KIT and DESY (Germany), 
INFN (Italy), 
SURF (Netherlands),
Polish WLCG (Poland),
IFIN-HH (Romania), 
PIC (Spain), CSCS (Switzerland), 
and GridPP (United Kingdom).
We are indebted to the communities behind the multiple open-source
software packages on which we depend.
Individual groups or members have received support from
Key Research Program of Frontier Sciences of CAS, CAS PIFI, CAS CCEPP, 
Minciencias (Colombia);
EPLANET, Marie Sk\l{}odowska-Curie Actions, ERC and NextGenerationEU (European Union);
A*MIDEX, ANR, IPhU and Labex P2IO, and R\'{e}gion Auvergne-Rh\^{o}ne-Alpes (France);
Alexander-von-Humboldt Foundation (Germany);
ICSC (Italy); 
Severo Ochoa and Mar\'ia de Maeztu Units of Excellence, GVA, XuntaGal, GENCAT, InTalent-Inditex and Prog.~Atracci\'on Talento CM (Spain);
SRC (Sweden);
the Leverhulme Trust, the Royal Society and UKRI (United Kingdom).

\emph{Data availability:} The data that support the findings of this article are openly available~\cite{LHCb-PAPER-2025-053-cds}.
\clearpage\newpage

\section*{End Matter}
\appendix
\section{Fit to the \Lc--\Lcbar mass distribution}
\label{sec:fit_to_LcLc}

\begin{figure}[!bp]
    \centering
   \subfloat{\includegraphics[width=0.5\textwidth]{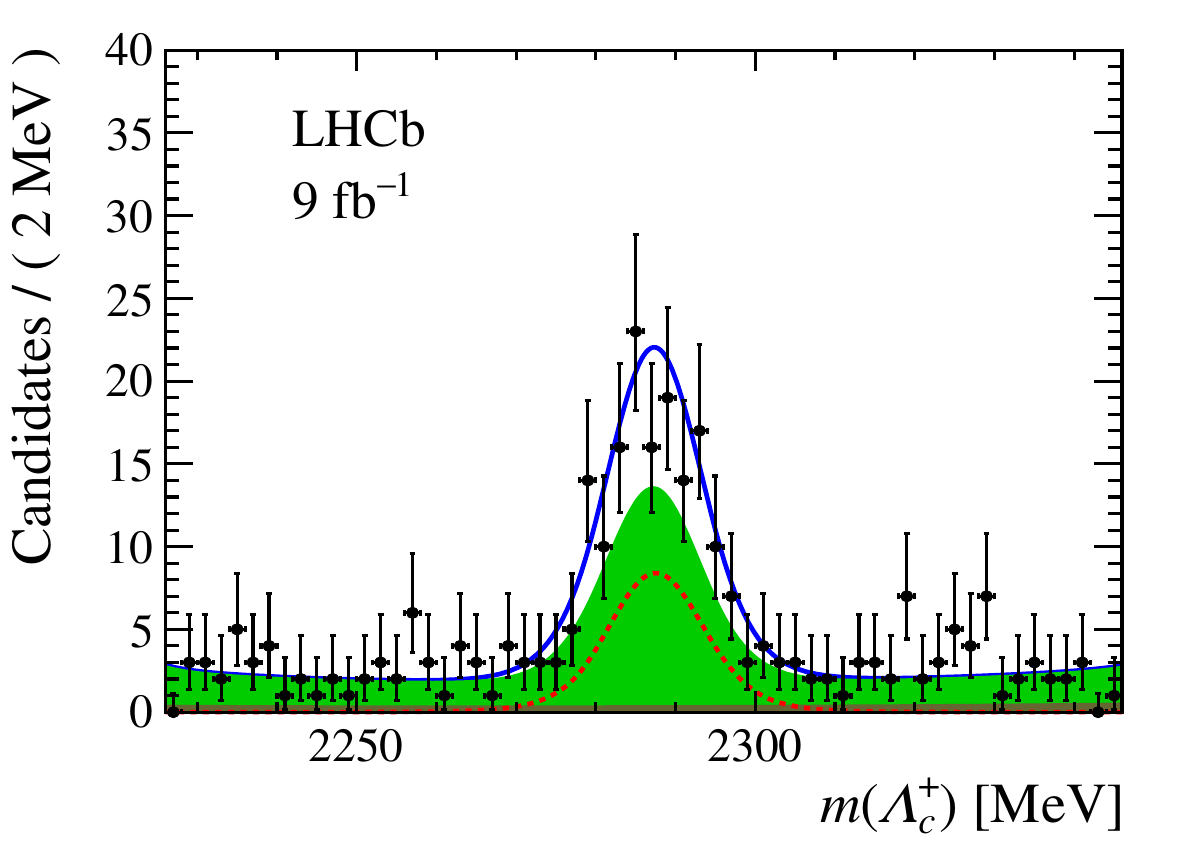}}
      \subfloat{\includegraphics[width=0.5\textwidth]{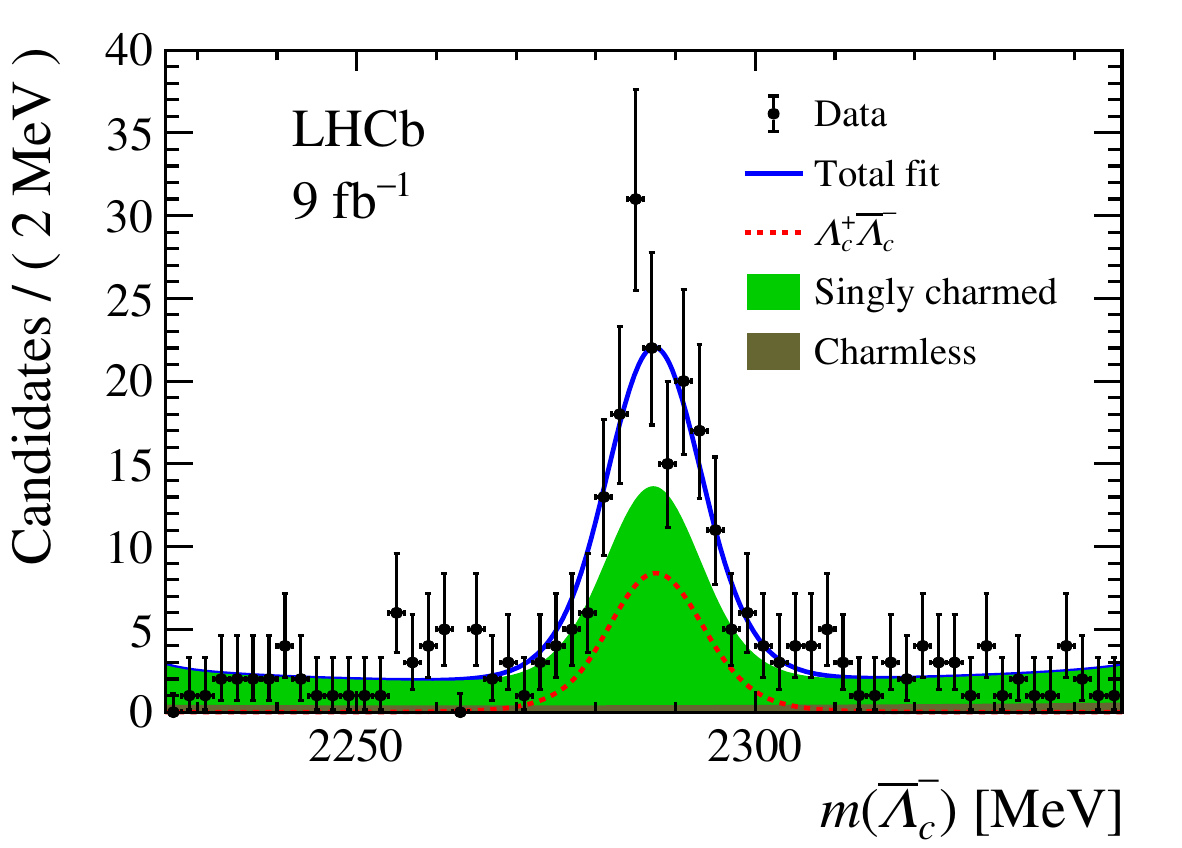}}
    \caption{Distributions of the (left) \Lc and (right) \Lcbar mass for the signal sample. Also shown is the projection of the two-dimensional fit (solid blue line) integrated over all $m(\Lc\Lcbar)$ bins, along with the contributions from doubly charmed (dashed red), singly charmed (filled green), and charmless (filled brown) decays.}
    \label{fig:LcLcbarMass}
\end{figure}

A two-dimensional fit model is constructed to describe the $[m(\Lc),m(\Lcbar)]$ distribution. In each dimension, the peaking component is modeled with a DSCB function $\mathcal{S}$, while the nonresonant background is described by an exponential function $\mathcal{B}$. The total probability density function is constructed from the product of all combinations of $\mathcal{S}$ and $\mathcal{B}$, as
\begin{eqnarray*}
\mathcal{P}(m(\Lc),m(\Lcbar)) &=& N_\text{DC}\cdot\left[\mathcal{S}(m(\Lc)) \mathcal{S}(m(\Lcbar))\right] \\
&+& N_\text{SC}\cdot\left[ \frac{1}{2}\mathcal{S}(m(\Lc)) \mathcal{B}(m(\Lcbar)) +\frac{1}{2} \mathcal{B}(m(\Lc)) \mathcal{S}(m(\Lcbar)) \right] \\
&+& N_\text{bkg}\cdot \left[ \mathcal{B}(m(\Lc)) \mathcal{B}(m(\Lcbar))\right],
\end{eqnarray*}
where $N_\text{DC}$, $N_\text{SC}$, and $N_\text{bkg}$ represent the yields of doubly charmed, singly charmed, and charmless background candidates, respectively. The factor $1/2$ arises from the charged-conjugate relationship between the \mbox{\decay{X}{\Lc\antiproton\Kp\pim}} and \mbox{\decay{\offsetoverline{X}}{\Lcbar\proton\Km\pip}} systems and has been verified using simulated samples. The fit is performed individually for each $m(\Lc\Lcbar)$ bin, and the corresponding $N_\text{DC}$ value and its uncertainty are determined. Figure~\ref{fig:LcLcbarMass} shows the $\Lc$ and $\Lcbar$ mass distributions with the fit results also included, where results from all individual $m(\Lc\Lcbar)$ bins are combined to improve visibility.

\section{Fit to the normalization decay modes}
\label{sec:fit_to_normalization}

\begin{figure}[!bp]
    \centering
   \subfloat{\includegraphics[width=0.5\textwidth]{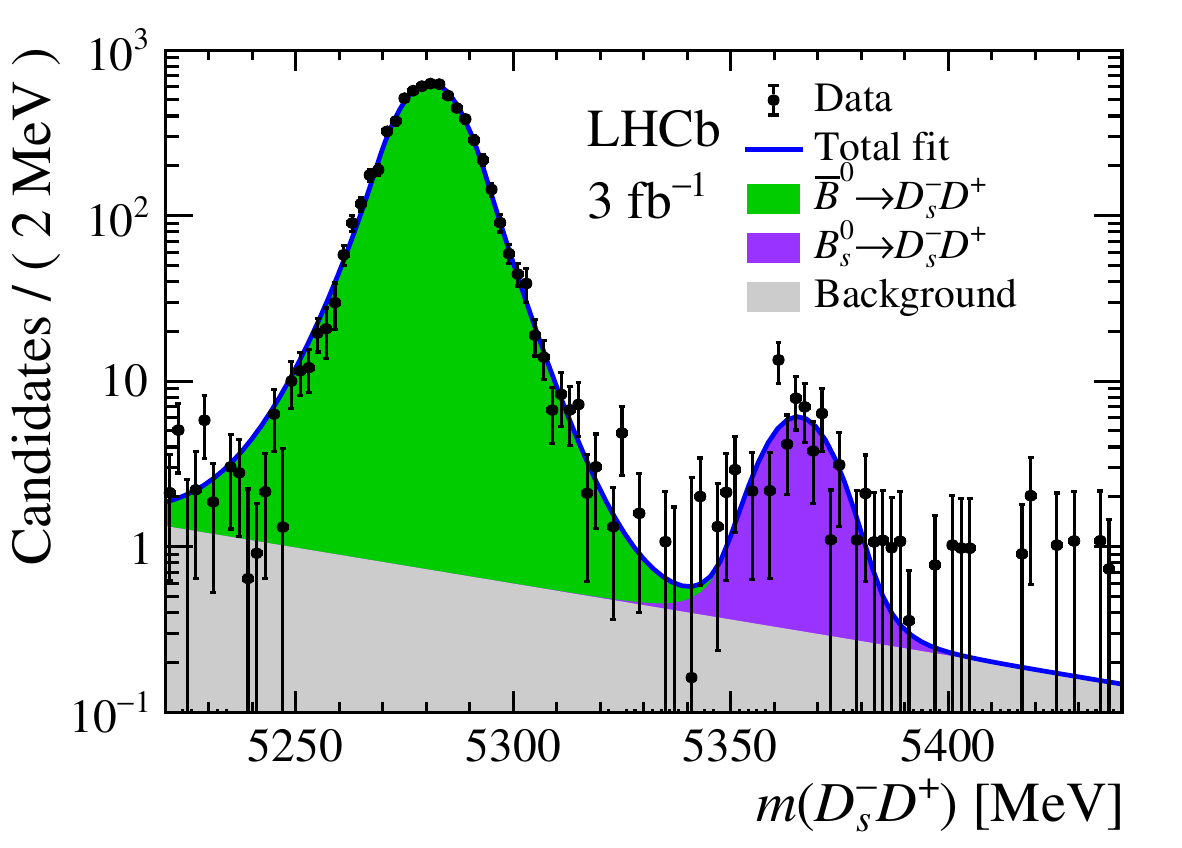}}
      \subfloat{\includegraphics[width=0.5\textwidth]{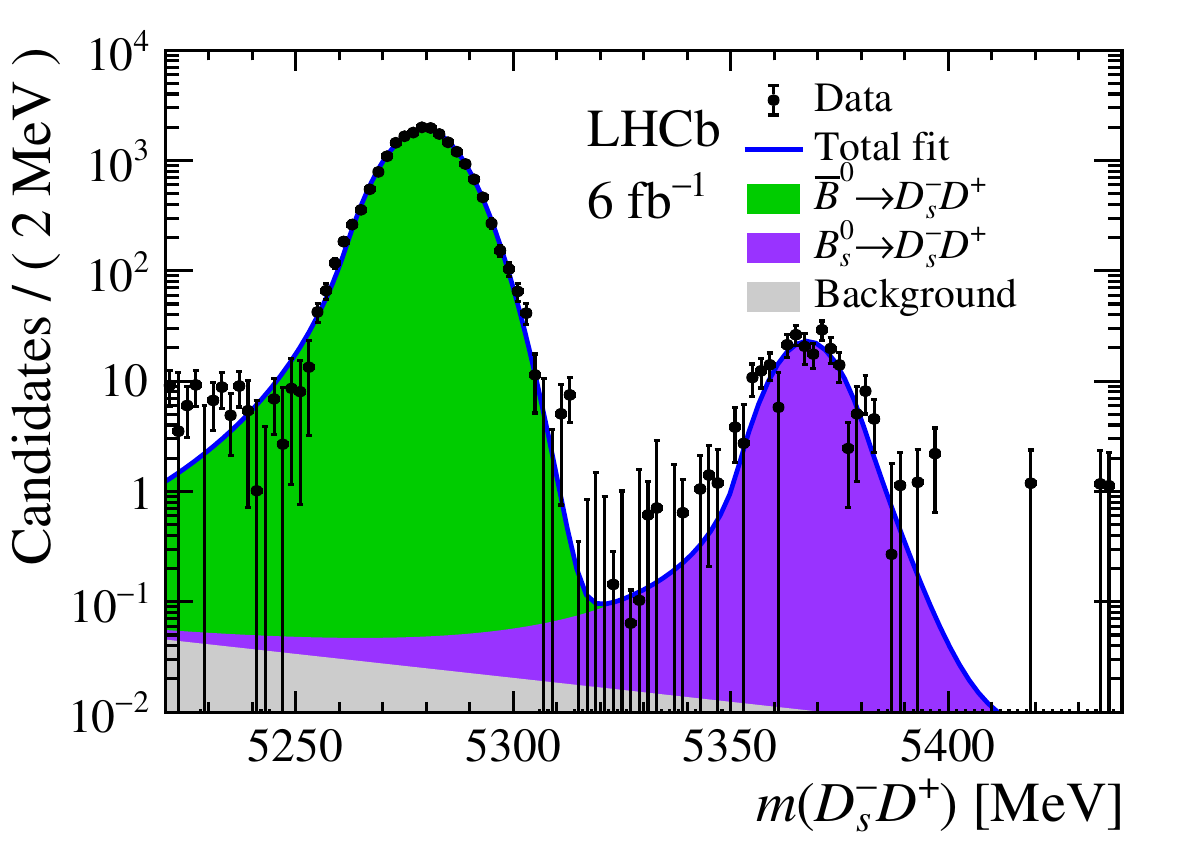}}
    \caption{Mass distribution of the $\Dsm\Dp$ system, with singly charmed and charmless backgrounds subtracted, shown for (left) Run~1 and (right) Run~2 data-taking periods. The fit projection (solid blue line) is shown together with the signal components from \BdDsD (filled green) and \BsDsD (filled purple) decays, and the background contribution (filled gray). 
    }
    \label{fig:NormMass_DsD}
\end{figure}

\begin{figure}[!bp]
    \centering
   \subfloat{\includegraphics[width=0.5\textwidth]{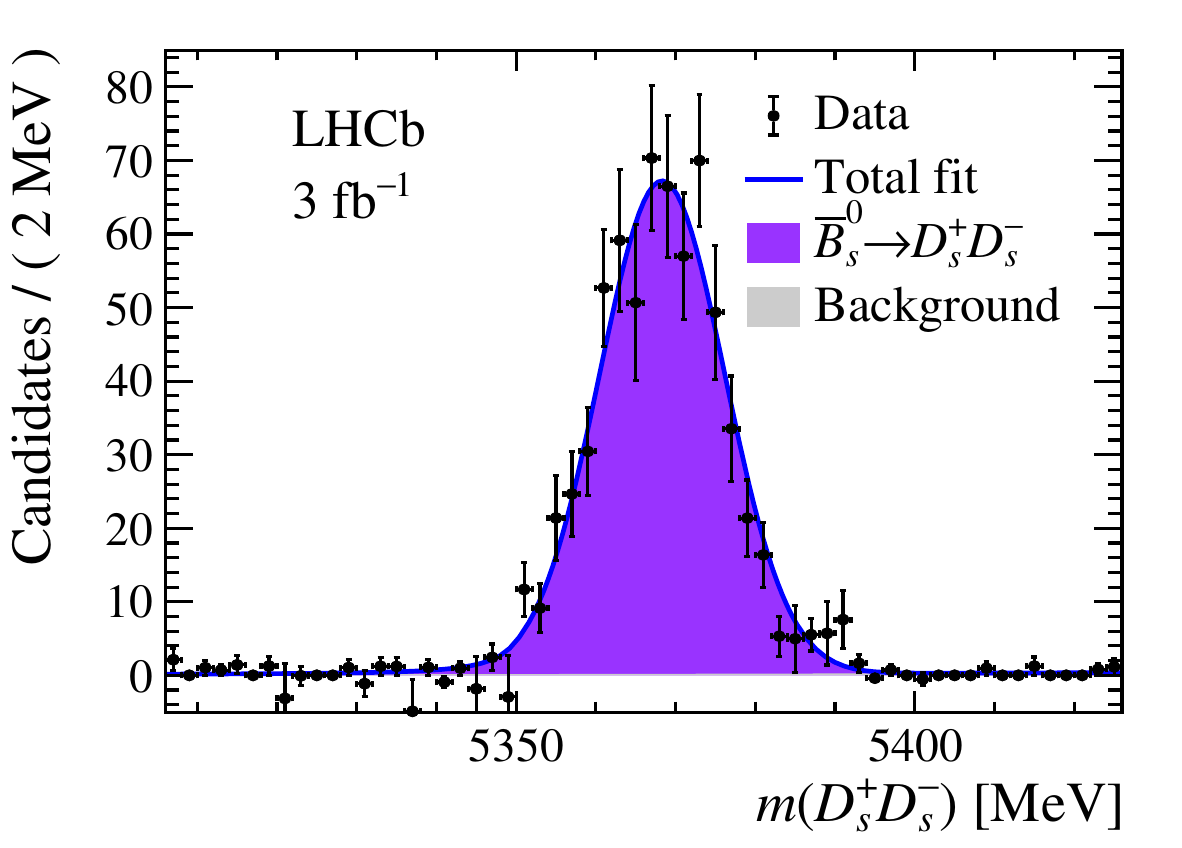}}
      \subfloat{\includegraphics[width=0.5\textwidth]{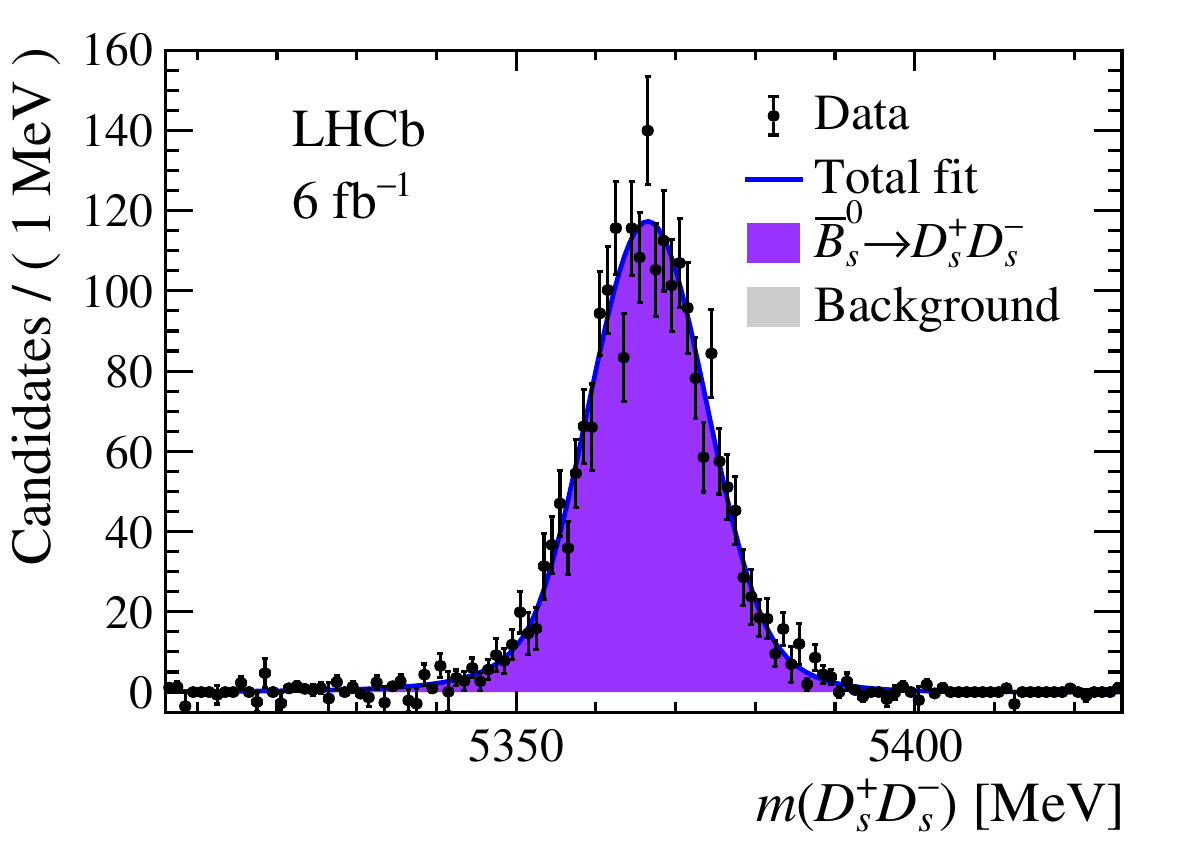}}
    \caption{Mass distribution of the $\Dsp\Dsm$ system, with singly charmed and charmless backgrounds subtracted, shown for (left) Run~1 and (right) Run~2 data-taking periods. The fit projection (solid blue line) is shown together with the \BsDsDs signal (filled purple) and the background (filled gray).}
    \label{fig:NormMass_DsDs}
\end{figure}

Figures~\ref{fig:NormMass_DsD} and~\ref{fig:NormMass_DsDs} show the $m(\Dsm\Dp)$ and $m(\Dsp\Dsm)$ distributions for the normalization decay modes, respectively, along with the fit projections.

\clearpage


\newpage
\addcontentsline{toc}{section}{References}
\bibliographystyle{LHCb}

\ifx\mcitethebibliography\mciteundefinedmacro
\PackageError{LHCb.bst}{mciteplus.sty has not been loaded}
{This bibstyle requires the use of the mciteplus package.}\fi
\providecommand{\href}[2]{#2}
\begin{mcitethebibliography}{10}
\mciteSetBstSublistMode{n}
\mciteSetBstMaxWidthForm{subitem}{\alph{mcitesubitemcount})}
\mciteSetBstSublistLabelBeginEnd{\mcitemaxwidthsubitemform\space}
{\relax}{\relax}

\bibitem{Cheng:2006nm}
H.-Y. Cheng, \ifthenelse{\boolean{articletitles}}{\emph{{Status of baryonic B decays}}, }{}\href{https://doi.org/10.1016/j.nuclphysbps.2006.09.004}{Nucl.\ Phys.\ {\bf B} Proc.\ Suppl.\  \textbf{163} (2007) 68}, \href{http://arxiv.org/abs/hep-ph/0607061}{{\normalfont\ttfamily arXiv:hep-ph/0607061}}\relax
\mciteBstWouldAddEndPuncttrue
\mciteSetBstMidEndSepPunct{\mcitedefaultmidpunct}
{\mcitedefaultendpunct}{\mcitedefaultseppunct}\relax
\EndOfBibitem
\bibitem{Huang:2021qld}
X.~Huang, Y.-K. Hsiao, J.~Wang, and L.~Sun, \ifthenelse{\boolean{articletitles}}{\emph{{Baryonic \B meson decays}}, }{}\href{https://doi.org/10.1155/2022/4343824}{Adv.\ High Energy Phys.\  \textbf{2022} (2022) 4343824}, \href{http://arxiv.org/abs/2109.02897}{{\normalfont\ttfamily arXiv:2109.02897}}\relax
\mciteBstWouldAddEndPuncttrue
\mciteSetBstMidEndSepPunct{\mcitedefaultmidpunct}
{\mcitedefaultendpunct}{\mcitedefaultseppunct}\relax
\EndOfBibitem
\bibitem{Cabibbo:1963yz}
N.~Cabibbo, \ifthenelse{\boolean{articletitles}}{\emph{{Unitary symmetry and leptonic decays}}, }{}\href{https://doi.org/10.1103/PhysRevLett.10.531}{Phys.\ Rev.\ Lett.\  \textbf{10} (1963) 531}\relax
\mciteBstWouldAddEndPuncttrue
\mciteSetBstMidEndSepPunct{\mcitedefaultmidpunct}
{\mcitedefaultendpunct}{\mcitedefaultseppunct}\relax
\EndOfBibitem
\bibitem{Kobayashi:1973fv}
M.~Kobayashi and T.~Maskawa, \ifthenelse{\boolean{articletitles}}{\emph{{\CP-violation in the renormalizable theory of weak interaction}}, }{}\href{https://doi.org/10.1143/PTP.49.652}{Prog.\ Theor.\ Phys.\  \textbf{49} (1973) 652}\relax
\mciteBstWouldAddEndPuncttrue
\mciteSetBstMidEndSepPunct{\mcitedefaultmidpunct}
{\mcitedefaultendpunct}{\mcitedefaultseppunct}\relax
\EndOfBibitem
\bibitem{Chen:2008pf}
C.-H. Chen, H.-Y. Cheng, and Y.-K. Hsiao, \ifthenelse{\boolean{articletitles}}{\emph{{Baryonic $D$ decay $D_s^{+}\to\proton\overline{n}$ and its implication}}, }{}\href{https://doi.org/10.1016/j.physletb.2008.04.033}{Phys.\ Lett.\  \textbf{B663} (2008) 326}, \href{http://arxiv.org/abs/0803.2910}{{\normalfont\ttfamily arXiv:0803.2910}}\relax
\mciteBstWouldAddEndPuncttrue
\mciteSetBstMidEndSepPunct{\mcitedefaultmidpunct}
{\mcitedefaultendpunct}{\mcitedefaultseppunct}\relax
\EndOfBibitem
\bibitem{BaBar:2014omp}
BaBar and Belle collaborations, A.~J. Bevan {\em et~al.}, \ifthenelse{\boolean{articletitles}}{\emph{{The physics of the \B factories}}, }{}\href{https://doi.org/10.1140/epjc/s10052-014-3026-9}{Eur.\ Phys.\ J.\  \textbf{C74} (2014) 3026}, \href{http://arxiv.org/abs/1406.6311}{{\normalfont\ttfamily arXiv:1406.6311}}\relax
\mciteBstWouldAddEndPuncttrue
\mciteSetBstMidEndSepPunct{\mcitedefaultmidpunct}
{\mcitedefaultendpunct}{\mcitedefaultseppunct}\relax
\EndOfBibitem
\bibitem{Chernyak:1990ag}
V.~L. Chernyak and I.~R. Zhitnitsky, \ifthenelse{\boolean{articletitles}}{\emph{{$B$-meson exclusive decays into baryons}}, }{}\href{https://doi.org/10.1016/0550-3213(90)90612-H}{Nucl.\ Phys.\  \textbf{B345} (1990) 137}\relax
\mciteBstWouldAddEndPuncttrue
\mciteSetBstMidEndSepPunct{\mcitedefaultmidpunct}
{\mcitedefaultendpunct}{\mcitedefaultseppunct}\relax
\EndOfBibitem
\bibitem{Cheng:2001tr}
H.-Y. Cheng and K.-C. Yang, \ifthenelse{\boolean{articletitles}}{\emph{{Charmless exclusive baryonic B decays}}, }{}\href{https://doi.org/10.1103/PhysRevD.66.014020}{Phys.\ Rev.\  \textbf{D66} (2002) 014020}, \href{http://arxiv.org/abs/hep-ph/0112245}{{\normalfont\ttfamily arXiv:hep-ph/0112245}}\relax
\mciteBstWouldAddEndPuncttrue
\mciteSetBstMidEndSepPunct{\mcitedefaultmidpunct}
{\mcitedefaultendpunct}{\mcitedefaultseppunct}\relax
\EndOfBibitem
\bibitem{Chang:2001jt}
C.-H.~V. Chang and W.-S. Hou, \ifthenelse{\boolean{articletitles}}{\emph{{B meson decays to baryons in the diquark model}}, }{}\href{https://doi.org/10.1007/s100520200894}{Eur.\ Phys.\ J.\  \textbf{C23} (2002) 691}, \href{http://arxiv.org/abs/hep-ph/0112219}{{\normalfont\ttfamily arXiv:hep-ph/0112219}}\relax
\mciteBstWouldAddEndPuncttrue
\mciteSetBstMidEndSepPunct{\mcitedefaultmidpunct}
{\mcitedefaultendpunct}{\mcitedefaultseppunct}\relax
\EndOfBibitem
\bibitem{Chua:2013zga}
C.-K. Chua, \ifthenelse{\boolean{articletitles}}{\emph{{Charmless two-body baryonic $B_{u,d,s}$ decays revisited}}, }{}\href{https://doi.org/10.1103/PhysRevD.89.056003}{Phys.\ Rev.\  \textbf{D89} (2014) 056003}, \href{http://arxiv.org/abs/1312.2335}{{\normalfont\ttfamily arXiv:1312.2335}}\relax
\mciteBstWouldAddEndPuncttrue
\mciteSetBstMidEndSepPunct{\mcitedefaultmidpunct}
{\mcitedefaultendpunct}{\mcitedefaultseppunct}\relax
\EndOfBibitem
\bibitem{He:2006vz}
X.-G. He, T.~Li, X.-Q. Li, and Y.-M. Wang, \ifthenelse{\boolean{articletitles}}{\emph{{Calculation of $\BF(\Bzb \to \Lc + \overline{p})$ in the perturbative QCD approach}}, }{}\href{https://doi.org/10.1103/PhysRevD.75.034011}{Phys.\ Rev.\  \textbf{D75} (2007) 034011}, \href{http://arxiv.org/abs/hep-ph/0607178}{{\normalfont\ttfamily arXiv:hep-ph/0607178}}\relax
\mciteBstWouldAddEndPuncttrue
\mciteSetBstMidEndSepPunct{\mcitedefaultmidpunct}
{\mcitedefaultendpunct}{\mcitedefaultseppunct}\relax
\EndOfBibitem
\bibitem{Cheng:2009yz}
H.-Y. Cheng, C.-K. Chua, and Y.-K. Hsiao, \ifthenelse{\boolean{articletitles}}{\emph{{Study of $\overline{B}\to\Lambda_c\overline{\Lambda}_{c}$ and $\overline{B}\to\Lambda_{c} \overline{\Lambda}_{c}\overline{K}$}}, }{}\href{https://doi.org/10.1103/PhysRevD.79.114004}{Phys.\ Rev.\  \textbf{D79} (2009) 114004}, \href{http://arxiv.org/abs/0902.4295}{{\normalfont\ttfamily arXiv:0902.4295}}\relax
\mciteBstWouldAddEndPuncttrue
\mciteSetBstMidEndSepPunct{\mcitedefaultmidpunct}
{\mcitedefaultendpunct}{\mcitedefaultseppunct}\relax
\EndOfBibitem
\bibitem{Geng:2025yna}
C.-Q. Geng, X.-N. Jin, and C.-W. Liu, \ifthenelse{\boolean{articletitles}}{\emph{{Large CP asymmetries from final-state interactions in charmful baryonic decays of $B^0{\rightarrow}{\Xi}_c^+{\Xi}_c^{-}$ and $B_s^{0}{\rightarrow}{\Lambda}_c^{+}{\Lambda}_c^{-}$}}, }{}\href{https://doi.org/10.1103/rgyz-jvgk}{Phys.\ Rev.\  \textbf{D112} (2025) 013001}, \href{http://arxiv.org/abs/2502.12770}{{\normalfont\ttfamily arXiv:2502.12770}}\relax
\mciteBstWouldAddEndPuncttrue
\mciteSetBstMidEndSepPunct{\mcitedefaultmidpunct}
{\mcitedefaultendpunct}{\mcitedefaultseppunct}\relax
\EndOfBibitem
\bibitem{LHCb-PAPER-2022-004}
LHCb collaboration, R.~Aaij {\em et~al.}, \ifthenelse{\boolean{articletitles}}{\emph{{Search for the rare hadronic decay $\Bs \to \proton\antiproton$}}, }{}\href{https://doi.org/10.1103/PhysRevD.108.012007}{Phys.\ Rev.\  \textbf{D108} (2023) 012007}, \href{http://arxiv.org/abs/2206.06673}{{\normalfont\ttfamily arXiv:2206.06673}}\relax
\mciteBstWouldAddEndPuncttrue
\mciteSetBstMidEndSepPunct{\mcitedefaultmidpunct}
{\mcitedefaultendpunct}{\mcitedefaultseppunct}\relax
\EndOfBibitem
\bibitem{Hsiao:2019wyd}
Y.-K. Hsiao, S.-Y. Tsai, C.-C. Lih, and E.~Rodrigues, \ifthenelse{\boolean{articletitles}}{\emph{{Testing the $W$-exchange mechanism with two-body baryonic $B$ decays}}, }{}\href{https://doi.org/10.1007/JHEP04(2020)035}{JHEP \textbf{04} (2020) 035}, \href{http://arxiv.org/abs/1906.01805}{{\normalfont\ttfamily arXiv:1906.01805}}\relax
\mciteBstWouldAddEndPuncttrue
\mciteSetBstMidEndSepPunct{\mcitedefaultmidpunct}
{\mcitedefaultendpunct}{\mcitedefaultseppunct}\relax
\EndOfBibitem
\bibitem{Hsiao:2023mud}
Y.-K. Hsiao, \ifthenelse{\boolean{articletitles}}{\emph{{Study of two-body doubly charmful baryonic $B$ decays with $SU(3)$ flavor symmetry}}, }{}\href{https://doi.org/10.1007/JHEP11(2023)117}{JHEP \textbf{11} (2023) 117}, \href{http://arxiv.org/abs/2309.16919}{{\normalfont\ttfamily arXiv:2309.16919}}\relax
\mciteBstWouldAddEndPuncttrue
\mciteSetBstMidEndSepPunct{\mcitedefaultmidpunct}
{\mcitedefaultendpunct}{\mcitedefaultseppunct}\relax
\EndOfBibitem
\bibitem{Geng:2024uxp}
C.-Q. Geng, X.-N. Jin, C.-W. Liu, and X.~Yu, \ifthenelse{\boolean{articletitles}}{\emph{{Hidden strangeness in meson weak decays to baryon pair}}, }{}\href{https://doi.org/10.1103/PhysRevD.110.113008}{Phys.\ Rev.\  \textbf{D110} (2024) 113008}, \href{http://arxiv.org/abs/2409.11374}{{\normalfont\ttfamily arXiv:2409.11374}}\relax
\mciteBstWouldAddEndPuncttrue
\mciteSetBstMidEndSepPunct{\mcitedefaultmidpunct}
{\mcitedefaultendpunct}{\mcitedefaultseppunct}\relax
\EndOfBibitem
\bibitem{Rui:2024xgc}
Z.~Rui, Z.-T. Zou, and Y.~Li, \ifthenelse{\boolean{articletitles}}{\emph{{Higher twist corrections to doubly-charmed baryonic B decays}}, }{}\href{https://doi.org/10.1007/JHEP12(2024)159}{JHEP \textbf{12} (2024) 159}, \href{http://arxiv.org/abs/2409.16113}{{\normalfont\ttfamily arXiv:2409.16113}}\relax
\mciteBstWouldAddEndPuncttrue
\mciteSetBstMidEndSepPunct{\mcitedefaultmidpunct}
{\mcitedefaultendpunct}{\mcitedefaultseppunct}\relax
\EndOfBibitem
\bibitem{PDG2024}
Particle Data Group, S.~Navas {\em et~al.}, \ifthenelse{\boolean{articletitles}}{\emph{{\href{http://pdg.lbl.gov/}{Review of particle physics}}}, }{}\href{https://doi.org/10.1103/PhysRevD.110.030001}{Phys.\ Rev.\  \textbf{D110} (2024) 030001}\relax
\mciteBstWouldAddEndPuncttrue
\mciteSetBstMidEndSepPunct{\mcitedefaultmidpunct}
{\mcitedefaultendpunct}{\mcitedefaultseppunct}\relax
\EndOfBibitem
\bibitem{Belle:2007lyc}
Belle collaboration, Y.~Uchida {\em et~al.}, \ifthenelse{\boolean{articletitles}}{\emph{{Search for $\overline{B}^{0}\rightarrow\Lambda_c^{+} \overline{\Lambda}_c^{-}$ decay at Belle}}, }{}\href{https://doi.org/10.1103/PhysRevD.77.051101}{Phys.\ Rev.\  \textbf{D77} (2008) 051101}, \href{http://arxiv.org/abs/0708.1105}{{\normalfont\ttfamily arXiv:0708.1105}}\relax
\mciteBstWouldAddEndPuncttrue
\mciteSetBstMidEndSepPunct{\mcitedefaultmidpunct}
{\mcitedefaultendpunct}{\mcitedefaultseppunct}\relax
\EndOfBibitem
\bibitem{LHCb-PAPER-2014-002}
LHCb collaboration, R.~Aaij {\em et~al.}, \ifthenelse{\boolean{articletitles}}{\emph{{Study of beauty hadron decays into pairs of charm hadrons}}, }{}\href{https://doi.org/10.1103/PhysRevLett.112.202001}{Phys.\ Rev.\ Lett.\  \textbf{112} (2014) 202001}, \href{http://arxiv.org/abs/1403.3606}{{\normalfont\ttfamily arXiv:1403.3606}}\relax
\mciteBstWouldAddEndPuncttrue
\mciteSetBstMidEndSepPunct{\mcitedefaultmidpunct}
{\mcitedefaultendpunct}{\mcitedefaultseppunct}\relax
\EndOfBibitem
\bibitem{LHCb-DP-2008-001}
LHCb collaboration, A.~A. Alves~Jr.\ {\em et~al.}, \ifthenelse{\boolean{articletitles}}{\emph{{The \lhcb detector at the LHC}}, }{}\href{https://doi.org/10.1088/1748-0221/3/08/S08005}{JINST \textbf{3} (2008) S08005}\relax
\mciteBstWouldAddEndPuncttrue
\mciteSetBstMidEndSepPunct{\mcitedefaultmidpunct}
{\mcitedefaultendpunct}{\mcitedefaultseppunct}\relax
\EndOfBibitem
\bibitem{LHCb-DP-2014-002}
LHCb collaboration, R.~Aaij {\em et~al.}, \ifthenelse{\boolean{articletitles}}{\emph{{LHCb detector performance}}, }{}\href{https://doi.org/10.1142/S0217751X15300227}{Int.\ J.\ Mod.\ Phys.\  \textbf{A30} (2015) 1530022}, \href{http://arxiv.org/abs/1412.6352}{{\normalfont\ttfamily arXiv:1412.6352}}\relax
\mciteBstWouldAddEndPuncttrue
\mciteSetBstMidEndSepPunct{\mcitedefaultmidpunct}
{\mcitedefaultendpunct}{\mcitedefaultseppunct}\relax
\EndOfBibitem
\bibitem{LHCb-DP-2012-004}
R.~Aaij {\em et~al.}, \ifthenelse{\boolean{articletitles}}{\emph{{The \lhcb trigger and its performance in 2011}}, }{}\href{https://doi.org/10.1088/1748-0221/8/04/P04022}{JINST \textbf{8} (2013) P04022}, \href{http://arxiv.org/abs/1211.3055}{{\normalfont\ttfamily arXiv:1211.3055}}\relax
\mciteBstWouldAddEndPuncttrue
\mciteSetBstMidEndSepPunct{\mcitedefaultmidpunct}
{\mcitedefaultendpunct}{\mcitedefaultseppunct}\relax
\EndOfBibitem
\bibitem{LHCb-DP-2019-001}
R.~Aaij {\em et~al.}, \ifthenelse{\boolean{articletitles}}{\emph{{Design and performance of the LHCb trigger and full real-time reconstruction in Run 2 of the LHC}}, }{}\href{https://doi.org/10.1088/1748-0221/14/04/P04013}{JINST \textbf{14} (2019) P04013}, \href{http://arxiv.org/abs/1812.10790}{{\normalfont\ttfamily arXiv:1812.10790}}\relax
\mciteBstWouldAddEndPuncttrue
\mciteSetBstMidEndSepPunct{\mcitedefaultmidpunct}
{\mcitedefaultendpunct}{\mcitedefaultseppunct}\relax
\EndOfBibitem
\bibitem{Sjostrand:2007gs}
T.~Sj\"{o}strand, S.~Mrenna, and P.~Skands, \ifthenelse{\boolean{articletitles}}{\emph{{A brief introduction to PYTHIA 8.1}}, }{}\href{https://doi.org/10.1016/j.cpc.2008.01.036}{Comput.\ Phys.\ Commun.\  \textbf{178} (2008) 852}, \href{http://arxiv.org/abs/0710.3820}{{\normalfont\ttfamily arXiv:0710.3820}}\relax
\mciteBstWouldAddEndPuncttrue
\mciteSetBstMidEndSepPunct{\mcitedefaultmidpunct}
{\mcitedefaultendpunct}{\mcitedefaultseppunct}\relax
\EndOfBibitem
\bibitem{LHCb-PROC-2010-056}
I.~Belyaev {\em et~al.}, \ifthenelse{\boolean{articletitles}}{\emph{{Handling of the generation of primary events in Gauss, the LHCb simulation framework}}, }{}\href{https://doi.org/10.1088/1742-6596/331/3/032047}{J.\ Phys.\ Conf.\ Ser.\  \textbf{331} (2011) 032047}\relax
\mciteBstWouldAddEndPuncttrue
\mciteSetBstMidEndSepPunct{\mcitedefaultmidpunct}
{\mcitedefaultendpunct}{\mcitedefaultseppunct}\relax
\EndOfBibitem
\bibitem{Lange:2001uf}
D.~J. Lange, \ifthenelse{\boolean{articletitles}}{\emph{{The EvtGen particle decay simulation package}}, }{}\href{https://doi.org/10.1016/S0168-9002(01)00089-4}{Nucl.\ Instrum.\ Meth.\  \textbf{A462} (2001) 152}\relax
\mciteBstWouldAddEndPuncttrue
\mciteSetBstMidEndSepPunct{\mcitedefaultmidpunct}
{\mcitedefaultendpunct}{\mcitedefaultseppunct}\relax
\EndOfBibitem
\bibitem{Photos-2}
N.~Davidson, T.~Przedzinski, and Z.~Was, \ifthenelse{\boolean{articletitles}}{\emph{{\photos interface in \cpp: Technical and physics documentation}}, }{}\href{https://doi.org/10.1016/j.cpc.2015.09.013}{Comput.\ Phys.\ Commun.\  \textbf{199} (2016) 86}, \href{http://arxiv.org/abs/1011.0937}{{\normalfont\ttfamily arXiv:1011.0937}}\relax
\mciteBstWouldAddEndPuncttrue
\mciteSetBstMidEndSepPunct{\mcitedefaultmidpunct}
{\mcitedefaultendpunct}{\mcitedefaultseppunct}\relax
\EndOfBibitem
\bibitem{Agostinelli:2002hh}
Geant4 collaboration, S.~Agostinelli {\em et~al.}, \ifthenelse{\boolean{articletitles}}{\emph{{Geant4: A simulation toolkit}}, }{}\href{https://doi.org/10.1016/S0168-9002(03)01368-8}{Nucl.\ Instrum.\ Meth.\  \textbf{A506} (2003) 250}\relax
\mciteBstWouldAddEndPuncttrue
\mciteSetBstMidEndSepPunct{\mcitedefaultmidpunct}
{\mcitedefaultendpunct}{\mcitedefaultseppunct}\relax
\EndOfBibitem
\bibitem{Allison:2006ve}
Geant4 collaboration, J.~Allison {\em et~al.}, \ifthenelse{\boolean{articletitles}}{\emph{{Geant4 developments and applications}}, }{}\href{https://doi.org/10.1109/TNS.2006.869826}{IEEE Trans.\ Nucl.\ Sci.\  \textbf{53} (2006) 270}\relax
\mciteBstWouldAddEndPuncttrue
\mciteSetBstMidEndSepPunct{\mcitedefaultmidpunct}
{\mcitedefaultendpunct}{\mcitedefaultseppunct}\relax
\EndOfBibitem
\bibitem{LHCb-PROC-2011-006}
M.~Clemencic {\em et~al.}, \ifthenelse{\boolean{articletitles}}{\emph{{The \lhcb simulation application, Gauss: Design, evolution and experience}}, }{}\href{https://doi.org/10.1088/1742-6596/331/3/032023}{J.\ Phys.\ Conf.\ Ser.\  \textbf{331} (2011) 032023}\relax
\mciteBstWouldAddEndPuncttrue
\mciteSetBstMidEndSepPunct{\mcitedefaultmidpunct}
{\mcitedefaultendpunct}{\mcitedefaultseppunct}\relax
\EndOfBibitem
\bibitem{LHCb-DP-2012-003}
M.~Adinolfi {\em et~al.}, \ifthenelse{\boolean{articletitles}}{\emph{{Performance of the \lhcb RICH detector at the LHC}}, }{}\href{https://doi.org/10.1140/epjc/s10052-013-2431-9}{Eur.\ Phys.\ J.\  \textbf{C73} (2013) 2431}, \href{http://arxiv.org/abs/1211.6759}{{\normalfont\ttfamily arXiv:1211.6759}}\relax
\mciteBstWouldAddEndPuncttrue
\mciteSetBstMidEndSepPunct{\mcitedefaultmidpunct}
{\mcitedefaultendpunct}{\mcitedefaultseppunct}\relax
\EndOfBibitem
\bibitem{LHCb-PUB-2016-021}
L.~Anderlini {\em et~al.}, \ifthenelse{\boolean{articletitles}}{\emph{{The PIDCalib package}}, }{} \href{http://cdsweb.cern.ch/search?p=LHCb-PUB-2016-021&f=reportnumber&action_search=Search&c=LHCb+Notes} {LHCb-PUB-2016-021}, 2016\relax
\mciteBstWouldAddEndPuncttrue
\mciteSetBstMidEndSepPunct{\mcitedefaultmidpunct}
{\mcitedefaultendpunct}{\mcitedefaultseppunct}\relax
\EndOfBibitem
\bibitem{Punzi:2003bu}
G.~Punzi, \ifthenelse{\boolean{articletitles}}{\emph{{Sensitivity of searches for new signals and its optimization}}, }{}eConf \textbf{C030908} (2003) MODT002, \href{http://arxiv.org/abs/physics/0308063}{{\normalfont\ttfamily arXiv:physics/0308063}}\relax
\mciteBstWouldAddEndPuncttrue
\mciteSetBstMidEndSepPunct{\mcitedefaultmidpunct}
{\mcitedefaultendpunct}{\mcitedefaultseppunct}\relax
\EndOfBibitem
\bibitem{Rogozhnikov:2016bdp}
A.~Rogozhnikov, \ifthenelse{\boolean{articletitles}}{\emph{{Reweighting with boosted decision trees}}, }{}\href{https://doi.org/10.1088/1742-6596/762/1/012036}{J.\ Phys.\ Conf.\ Ser.\  \textbf{762} (2016) 012036}, \href{http://arxiv.org/abs/1608.05806}{{\normalfont\ttfamily arXiv:1608.05806}}, \url{https://github.com/arogozhnikov/hep_ml}\relax
\mciteBstWouldAddEndPuncttrue
\mciteSetBstMidEndSepPunct{\mcitedefaultmidpunct}
{\mcitedefaultendpunct}{\mcitedefaultseppunct}\relax
\EndOfBibitem
\bibitem{Pivk:2004ty}
M.~Pivk and F.~R. Le~Diberder, \ifthenelse{\boolean{articletitles}}{\emph{{sPlot: A statistical tool to unfold data distributions}}, }{}\href{https://doi.org/10.1016/j.nima.2005.08.106}{Nucl.\ Instrum.\ Meth.\  \textbf{A555} (2005) 356}, \href{http://arxiv.org/abs/physics/0402083}{{\normalfont\ttfamily arXiv:physics/0402083}}\relax
\mciteBstWouldAddEndPuncttrue
\mciteSetBstMidEndSepPunct{\mcitedefaultmidpunct}
{\mcitedefaultendpunct}{\mcitedefaultseppunct}\relax
\EndOfBibitem
\bibitem{Skwarnicki:1986xj}
T.~Skwarnicki, {\em {A study of the radiative cascade transitions between the Upsilon-prime and Upsilon resonances}}, PhD thesis, Institute of Nuclear Physics, Krakow, 1986, {\href{http://inspirehep.net/record/230779/}{DESY-F31-86-02}}\relax
\mciteBstWouldAddEndPuncttrue
\mciteSetBstMidEndSepPunct{\mcitedefaultmidpunct}
{\mcitedefaultendpunct}{\mcitedefaultseppunct}\relax
\EndOfBibitem
\bibitem{Cranmer:2000du}
K.~S. Cranmer, \ifthenelse{\boolean{articletitles}}{\emph{{Kernel estimation in high-energy physics}}, }{}\href{https://doi.org/10.1016/S0010-4655(00)00243-5}{Comput.\ Phys.\ Commun.\  \textbf{136} (2001) 198}, \href{http://arxiv.org/abs/hep-ex/0011057}{{\normalfont\ttfamily arXiv:hep-ex/0011057}}\relax
\mciteBstWouldAddEndPuncttrue
\mciteSetBstMidEndSepPunct{\mcitedefaultmidpunct}
{\mcitedefaultendpunct}{\mcitedefaultseppunct}\relax
\EndOfBibitem
\bibitem{Santos:2013gra}
D.~Mart{\'\i}nez~Santos and F.~Dupertuis, \ifthenelse{\boolean{articletitles}}{\emph{{Mass distributions marginalized over per-event errors}}, }{}\href{https://doi.org/10.1016/j.nima.2014.06.081}{Nucl.\ Instrum.\ Meth.\  \textbf{A764} (2014) 150}, \href{http://arxiv.org/abs/1312.5000}{{\normalfont\ttfamily arXiv:1312.5000}}\relax
\mciteBstWouldAddEndPuncttrue
\mciteSetBstMidEndSepPunct{\mcitedefaultmidpunct}
{\mcitedefaultendpunct}{\mcitedefaultseppunct}\relax
\EndOfBibitem
\bibitem{rice2006mat}
J.~A. Rice, {\em Mathematical statistics and data analysis}, Belmont, CA: Duxbury Press., third~ed., 2006\relax
\mciteBstWouldAddEndPuncttrue
\mciteSetBstMidEndSepPunct{\mcitedefaultmidpunct}
{\mcitedefaultendpunct}{\mcitedefaultseppunct}\relax
\EndOfBibitem
\bibitem{LHCb-DP-2013-002}
LHCb collaboration, R.~Aaij {\em et~al.}, \ifthenelse{\boolean{articletitles}}{\emph{{Measurement of the track reconstruction efficiency at LHCb}}, }{}\href{https://doi.org/10.1088/1748-0221/10/02/P02007}{JINST \textbf{10} (2015) P02007}, \href{http://arxiv.org/abs/1408.1251}{{\normalfont\ttfamily arXiv:1408.1251}}\relax
\mciteBstWouldAddEndPuncttrue
\mciteSetBstMidEndSepPunct{\mcitedefaultmidpunct}
{\mcitedefaultendpunct}{\mcitedefaultseppunct}\relax
\EndOfBibitem
\bibitem{LHCb-PAPER-2022-002}
LHCb collaboration, R.~Aaij {\em et~al.}, \ifthenelse{\boolean{articletitles}}{\emph{{Amplitude analysis of the $\Lc \to \proton \Km\pip$ decay and $\Lc$ baryon polarization measurement in semileptonic beauty hadron decays}}, }{}\href{https://doi.org/10.1103/PhysRevD.108.012023}{Phys.\ Rev.\  \textbf{D108} (2023) 012023}, \href{http://arxiv.org/abs/2208.03262}{{\normalfont\ttfamily arXiv:2208.03262}}\relax
\mciteBstWouldAddEndPuncttrue
\mciteSetBstMidEndSepPunct{\mcitedefaultmidpunct}
{\mcitedefaultendpunct}{\mcitedefaultseppunct}\relax
\EndOfBibitem
\bibitem{LHCb-TDR-012}
LHCb collaboration, \ifthenelse{\boolean{articletitles}}{\emph{{Framework TDR for the LHCb Upgrade: Technical Design Report}}, }{} \href{http://cdsweb.cern.ch/search?p=CERN-LHCC-2012-007&f=reportnumber&action_search=Search&c=LHCb} {CERN-LHCC-2012-007}, 2012\relax
\mciteBstWouldAddEndPuncttrue
\mciteSetBstMidEndSepPunct{\mcitedefaultmidpunct}
{\mcitedefaultendpunct}{\mcitedefaultseppunct}\relax
\EndOfBibitem
\bibitem{LHCb-DP-2022-002}
LHCb collaboration, R.~Aaij {\em et~al.}, \ifthenelse{\boolean{articletitles}}{\emph{{The LHCb Upgrade I}}, }{}\href{https://doi.org/10.1088/1748-0221/19/05/P05065}{{JINST} \textbf{19} (2024) P05065}, \href{http://arxiv.org/abs/2305.10515}{{\normalfont\ttfamily arXiv:2305.10515}}\relax
\mciteBstWouldAddEndPuncttrue
\mciteSetBstMidEndSepPunct{\mcitedefaultmidpunct}
{\mcitedefaultendpunct}{\mcitedefaultseppunct}\relax
\EndOfBibitem
\bibitem{LHCb-PAPER-2025-053-cds}
\url{https://cds.cern.ch/record/2949754}\relax
\mciteBstWouldAddEndPuncttrue
\mciteSetBstMidEndSepPunct{\mcitedefaultmidpunct}
{\mcitedefaultendpunct}{\mcitedefaultseppunct}\relax
\EndOfBibitem
\end{mcitethebibliography}
\ifx\mcitethebibliography\mciteundefinedmacro
\PackageError{LHCb.bst}{mciteplus.sty has not been loaded}
{This bibstyle requires the use of the mciteplus package.}\fi
\providecommand{\href}[2]{#2}

\newpage
\centerline
{\large\bf LHCb collaboration}
\begin
{flushleft}
\small
R.~Aaij$^{38}$\lhcborcid{0000-0003-0533-1952},
A.S.W.~Abdelmotteleb$^{58}$\lhcborcid{0000-0001-7905-0542},
C.~Abellan~Beteta$^{52}$\lhcborcid{0009-0009-0869-6798},
F.~Abudin{\'e}n$^{58}$\lhcborcid{0000-0002-6737-3528},
T.~Ackernley$^{62}$\lhcborcid{0000-0002-5951-3498},
A. A. ~Adefisoye$^{70}$\lhcborcid{0000-0003-2448-1550},
B.~Adeva$^{48}$\lhcborcid{0000-0001-9756-3712},
M.~Adinolfi$^{56}$\lhcborcid{0000-0002-1326-1264},
P.~Adlarson$^{86}$\lhcborcid{0000-0001-6280-3851},
C.~Agapopoulou$^{14}$\lhcborcid{0000-0002-2368-0147}
% ...

{\footnotesize \it

% ...
$^{14}$Universit{\'e} Paris-Saclay, CNRS/IN2P3, IJCLab, Orsay, France\\
% ...
$^{38}$Nikhef National Institute for Subatomic Physics, Amsterdam, Netherlands\\
% ...
$^{48}$Instituto Galego de F{\'\i}sica de Altas Enerx{\'\i}as (IGFAE), Universidade de Santiago de Compostela, Santiago de Compostela, Spain\\
% ...
$^{52}$Physik-Institut, Universit{\"a}t Z{\"u}rich, Z{\"u}rich, Switzerland\\
% ...
$^{56}$H.H. Wills Physics Laboratory, University of Bristol, Bristol, United Kingdom\\
% ...
$^{58}$Department of Physics, University of Warwick, Coventry, United Kingdom\\
% ...
$^{61}$School of Physics and Astronomy, University of Glasgow, Glasgow, United Kingdom\\
$^{62}$Oliver Lodge Laboratory, University of Liverpool, Liverpool, United Kingdom\\
% ...
$^{70}$Syracuse University, Syracuse, NY, United States\\
% ...
$^{86}$Department of Physics and Astronomy, Uppsala University, Uppsala, Sweden, associated to $^{61}$\\
% ...
\medskip
}
\end{flushleft}

\end{document}